\documentclass[sigconf]{acmart}

\AtBeginDocument{%
  }

\copyrightyear{2025} 
\acmYear{2025}
\setcopyright{cc}
\setcctype{by}
\acmConference[ASIA CCS '25]{ACM Asia Conference on Computer and Communications Security}{August 25--29, 2025}{Hanoi, Vietnam}
\acmBooktitle{ACM Asia Conference on Computer and Communications Security (ASIA CCS '25), August 25--29, 2025, Hanoi, Vietnam}
\acmDOI{10.1145/3708821.3733872}
\acmISBN{979-8-4007-1410-8/2025/08}

\usepackage{pifont}
\usepackage{enumitem}
\usepackage{graphicx}
\usepackage{multirow}
\usepackage{tikz}
\usepackage{booktabs}
\usepackage{pifont}
\usepackage{soul}
\usepackage{array}
\usepackage{subcaption}
\usepackage{balance}
\usepackage{xcolor,colortbl}
\usepackage{algorithm}
\usepackage{algpseudocode}
\usepackage{listings}
\usepackage{xcolor}

\newcolumntype{P}[1]{>{\centering\arraybackslash}p{#1}}
\newcommand{\xmark}{\ding{55}}%

\newcommand*\circled[1]{\tikz[baseline=(char.base)]{
            \node[shape=circle,draw,inner sep=0.7pt] (char) {#1};}}

\usetikzlibrary{calc}

\makeatletter
\newif\if@anonymize

\@anonymizetrue    

\if@anonymize
  \newcommand{\highlight@DoHighlight}{
    \fill [outer sep = -15pt, inner sep = 0pt, color=black]
          ($(begin highlight)+(0,8pt)$) rectangle ($(end
highlight)+(0,-3pt)$) ;
  }

  \newcommand{\highlight@BeginHighlight}{
    \coordinate (begin highlight) at (0,0) ;
  }

  \newcommand{\highlight@EndHighlight}{
    \coordinate (end highlight) at (0,0) ;
  }

  \newdimen\highlight@previous
  \newdimen\highlight@current
  \newlength{\item@width}

  \DeclareRobustCommand*\anonymize{%
    \SOUL@setup
    \def\SOUL@preamble{%
      \begin{tikzpicture}[overlay, remember picture]
        \highlight@BeginHighlight
        \highlight@EndHighlight
      \end{tikzpicture}%
    }%
    \def\SOUL@postamble{%
      \begin{tikzpicture}[overlay, remember picture]
        \highlight@EndHighlight
        \highlight@DoHighlight
      \end{tikzpicture}%
    }%
    \def\SOUL@everyhyphen{%
      \discretionary{%
        \SOUL@setkern\SOUL@hyphkern
        \SOUL@sethyphenchar
        \tikz[overlay, remember picture] \highlight@EndHighlight ;%
      }{%
      }{%
        \SOUL@setkern\SOUL@charkern
      }%
    }%
    \def\SOUL@everyexhyphen##1{%
      \SOUL@setkern\SOUL@hyphkern
      \settowidth{\item@width}{##1}%
      \makebox[\item@width]{}%
      \discretionary{%
        \tikz[overlay, remember picture] \highlight@EndHighlight ;%
      }{%
      }{%
        \SOUL@setkern\SOUL@charkern
      }%
    }%
    \def\SOUL@everysyllable{%
      \begin{tikzpicture}[overlay, remember picture]
        \path let \p0 = (begin highlight), \p1 = (0,0) in \pgfextra
          \global\highlight@previous=\y0
          \global\highlight@current =\y1
        \endpgfextra (0,0) ;
        \ifdim\highlight@current < \highlight@previous
          \highlight@DoHighlight
          \highlight@BeginHighlight
        \fi
      \end{tikzpicture}%
      \settowidth{\item@width}{\the\SOUL@syllable}%
      \makebox[\item@width]{}%
      \tikz[overlay, remember picture] \highlight@EndHighlight ;%
    }%
    \SOUL@
  }
\else
  \newcommand{\anonymize}[1]{#1}
\fi
\makeatother

\newcommand{\ie}{\textit{i.e., }}
\newcommand{\eg}{\textit{e.g., }}

\newcommand{\cf}{\textit{cf. }}

\def \1{\textit{(i)}}
\def \2{\textit{(ii)}}
\def \3{\textit{(iii)}}
\def \4{\textit{(iv)}}
\def \5{\textit{(v)}}
\def \6{\textit{(vi)}}
\def \7{\textit{(vii)}}
\def \8{\textit{(viii)}}
\def \9{\textit{(ix)}}

\begin{document}

\title{QUIC-Exfil: Exploiting QUIC's Server Preferred Address Feature to Perform Data Exfiltration Attacks}


\author{Thomas Grübl}
\affiliation{%
  \institution{Communication Systems Group CSG, Department of Informatics IfI, University of Zurich UZH}
  \streetaddress{}
  \city{Zürich}
  \postcode{CH-8050}
  \country{Switzerland}}
\email{gruebl@ifi.uzh.ch}

\author{Weijie Niu}
\affiliation{%
 \institution{Communication Systems Group CSG, Department of Informatics IfI, University of Zurich UZH}
  \streetaddress{}
  \city{Zürich}
  \postcode{CH-8050}
  \country{Switzerland}}
\email{niu@ifi.uzh.ch}

\author{Jan von der Assen}
\affiliation{%
  \institution{Communication Systems Group CSG, Department of Informatics IfI, University of Zurich UZH}
  \streetaddress{}
  \city{Zürich}
  \postcode{CH-8050}
  \country{Switzerland}}
\email{vonderassen@ifi.uzh.ch}

\author{Burkhard Stiller}
\affiliation{%
  \institution{Communication Systems Group CSG, Department of Informatics IfI, University of Zurich UZH}
  \streetaddress{}
  \city{Zürich}
  \postcode{CH-8050}
  \country{Switzerland}}
\email{stiller@ifi.uzh.ch}

\renewcommand{\shortauthors}{Grübl et al.}

\begin{abstract}
The QUIC protocol is now widely adopted by major tech companies and accounts for a significant fraction of today's Internet traffic. QUIC's multiplexing capabilities, encrypted headers, dynamic IP address changes, and encrypted parameter negotiations make the protocol not only more efficient, secure, and censorship-resistant, but also practically unmanageable by firewalls. This opens up doors for attackers that may exploit certain traits of the QUIC protocol to perform targeted attacks, such as data exfiltration attacks. Whereas existing data exfiltration techniques, such as TLS and DNS-based exfiltration, can be detected on a firewall level, QUIC-based data exfiltration is more difficult to detect, since changes in IP addresses and ports are inherent to the protocol's normal behaviour.

To show the feasibility of a QUIC-based data exfiltration attack, we first introduce a novel method which leverages the server preferred address feature of the QUIC protocol and, thus, allows an attacker to exfiltrate sensitive data from an infected machine to a malicious server, disguised as a server-side connection migration. The attack is implemented in the form of a proof of concept tool in Rust. We evaluated the performance of five anomaly detection classifiers --- Random Forest, Multi-Layer Perceptron, Support Vector Machine, Autoencoder, and Isolation Forest --- trained on datasets collected from three distinct network traffic scenarios. The classifiers were trained on $\sim$~700K benign and malicious QUIC packets and 786  connection migration events, but were unable to effectively detect the data exfiltration attempts. Furthermore, post-analysis of the traffic captures did not reveal any identifiable fingerprint. As part of our evaluation, we also interviewed five leading firewall vendors and found that, as of today, no major firewall vendor implements functionality capable of distinguishing between benign and malicious QUIC connection migrations.
\end{abstract}




\maketitle


\section{Introduction}

The QUIC protocol, originally developed by \cite{roskind2012}, is an IETF standardized, connection-oriented, UDP-based transport protocol that serves as a replacement to TLS over TCP \cite{langley2017}. Its main benefits include One Round Trip Time (1-RTT) handshakes, Zero Round Trip Time (0-RTT) connection re-establishments when prior communication has taken place, seamless network path migration, and state-of-the-art security \cite{rfc9000}. As of today, QUIC has been widely deployed by major tech companies such as Google, Meta, Apple, and Cloudflare. The popularity of QUIC has led to some application-layer protocols adopting it as their main transport protocol, such as DNS-over-QUIC \cite{rfc9250} or HTTP/3 \cite{rfc9114}. QUIC plays a fundamental role within HTTP/3 \cite{rfc9114}, making it ubiquitous in today's Internet traffic. In 2023, HTTP/3 already accounted for around 30\% of HTTP requests, with adoption expected to continue growing as more web servers and browsers integrate QUIC support \cite{cloudflarehttp3}.

QUIC packets are encapsulated in UDP datagrams. Each packet has a QUIC header that contains details such as the flags indicating a certain packet type and the Connection ID (CID). The CID is a randomly generated, variable-length identifier that endpoints use to demultiplex network traffic sent between them \cite{rfc9000}. Since the QUIC protocol encrypts the entire communication, including a significant part of the handshake, it is inherently difficult to analyze the underlying traffic. Although QUIC was primarily designed for performance improvements,  QUIC's properties also enhance privacy and make it more resilient to network interference through features such as the frequent rotation of CIDs and the concurrent transmission of packets over multiple streams \cite{martini2019quic}.

While these features improve the security and privacy for end-users, they pose challenges for firewall vendors. The general consensus among firewall vendors and administrators appears to be that QUIC-based traffic should be blocked because it renders Deep Packet Inspection (DPI) and Intrusion Detection/Prevention Systems (IDS/IPS) useless (\cf Section \ref{sec:evaluation}). Typically, the communication between client and server is then forced to fall back to TLS over TCP, allowing enterprise firewalls to perform HTTPS DPI \cite{ciscoevi}. As of now, firewall vendors do not provide any specifics on how enterprise-level firewalls can filter QUIC traffic effectively.

Some inherent traits of the QUIC protocol, such as the connection migration capability and the encrypted handshake, make QUIC-based data exfiltration techniques arguably more devious than TLS or DNS-based data exfiltration because middleboxes are not able to differentiate between benign and spoofed connection migration attempts. It is, therefore, crucial to understand how the underlying properties of the QUIC protocol can impact the success of data exfiltration attacks.

The goal of data exfiltration is to covertly extract sensitive data from an infected host. There exist many methods to exfiltrate data --- depending on the underlying protocol used, some are more effective than others. According to the MITRE ATT\&CK framework \cite{mitre}, data exfiltration attacks can be broadly divided into network-based and physical exfiltration methods. The former leverages widely-used network protocols, such as HTTP, HTTPS, or DNS, to send data to a target device owned by a malicious actor. As part of the latter, adversaries may use removable storage, such as external hard drives or USB drives, to exfiltrate data. Effectiveness in this context can be seen as a combination of data throughput and covertness of the method. The severity of data exfiltration attacks becomes apparent when considering the financial impact of data breaches. The average cost of data breaches has been consistently rising over the last years, with current estimations reaching a global all-time high of 4.88 million USD per data breach. The top three industries to experience the highest data breach costs are healthcare, financial, and industrial~\cite{ibm}.

In this paper, we specifically focus on QUIC's server-side connection migration feature. We show that an attack can be designed that replicates the behavior of this feature to covertly exfiltrate data to a target server. The attack can be leveraged by adversaries that have already infected a host machine. From the perspective of a middlebox, the exfiltration payload packets are not distinguishable from packets that originate from a benign server-side connection migration. While we initially explored classifiers as a potential defense strategy to detect QUIC-based data exfiltration traffic, our findings indicate that machine learning-based classifiers are ineffective in reliably distinguishing the attack from legitimate traffic. To the best of our knowledge, this paper provides the first analysis of QUIC-based data exfiltration attacks and potential mitigations. The \textbf{contributions} of the paper can be summarized as follows:

\begin{itemize}[label={$\checkmark$}]
    \item We develop a QUIC-based data exfiltration method that exploits the QUIC connection migration feature by mimicking a server-side path migration procedure to exfiltrate sensitive data from an infected host machine. The method does not require any changes to the QUIC client or server applications and is designed based on the version-independent properties of IETF QUIC.
    \item We implement a PoC prototype in Rust, which comprises a packet sniffer and a custom QUIC parser and mimics benign packet features such as payload length, entropy, time differences between outgoing packets as well as the server-side connection migration packets.
    \item We show that due to the limited number of visible features available in QUIC traffic, it is difficult to differentiate between benign and malicious QUIC traffic. Our custom detectors --- Random Forest, Multi-Layer Perceptron, Support Vector Machine, Autoencoder, and Isolation Forest --- were trained on over 700K QUIC packets and 786 server-side connection migration events, collected across three distinct network traffic scenarios.
    \item In addition, we conducted unstructured interviews with five leading firewall vendors to gain insights into the current state of their QUIC traffic filtering capabilities.
\end{itemize}

First, we begin with the background (Section \ref{sec:background}) and the problem statement (Section \ref{sec:problem_statement}). Section \ref{sec:methodology} presents the data exfiltration method and the threat model. Section \ref{sec:implementation} describes the PoC implementation written in Rust. Section \ref{sec:evaluation} presents the experimental setup, the anomaly detection results and the survey of leading firewall vendors, followed by a brief discussion of additional mitigation strategies in Section \ref{sec:discussion}, the related work in Section \ref{sec:related_work}, and the conclusion in Section \ref{sec:conclusion}.

\section{Background}\label{sec:background}

This section briefly introduces the main characteristics of the QUIC protocol, focusing on its header structures and the connection migration features for both client-side and server-side migrations since those are the most relevant aspects for the attack. The description of the packet headers mainly relies on the version-independent properties of QUIC, defined in RFC 8999 \cite{rfc8999}, and connection migrations are described based on RFC 9000 \cite{rfc9000}.

\subsection{A QUIC Overview}

QUIC packets generally come in two different forms: long header and short header packets. Long header packets are sent as part of the handshake messages before 1-RTT keys are established. Thereafter, the QUIC endpoints switch to sending short header packets \cite{rfc9000}. Long header packets include version-specific bits, the version number, CIDs for both source and destination as well as their respective lengths, other version-dependent data, and the payload itself. In comparison, short header packets only contain version-specific bits, the Destination Connection ID (DCID), version-dependent data, and the payload, meaning that the length of the CID cannot be determined from a short header packet alone. Without observing the initial long header packets, the necessary context for understanding short header packets, \eg which connection they belong to, cannot be established. All QUIC endpoints, therefore, need to keep track of the CIDs in use \cite{rfc8999}.


\subsection{Connection Migration}

\begin{figure}[!b]
\centering
\includegraphics[width=2.4in]{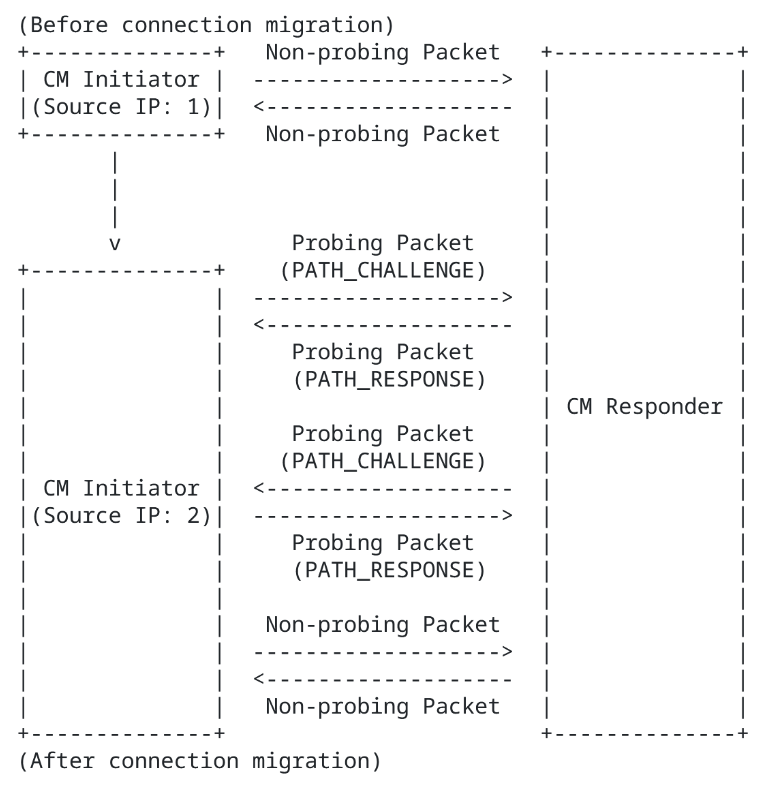}
\caption{Client-side QUIC Connection Migration \cite{tan2020}}
\label{fig:conn_migration}
\end{figure}

Compared to traditional TCP and UDP connections, where each connection is uniquely identifiable by its 5-tuple (source IP address, destination IP address, source port, destination port, protocol), a QUIC connection may change its underlying 5-tuple without establishing a new connection using a handshake. This is referred to as connection migration or (network) path migration. The primary benefit of QUIC connection migrations lies in the reduction of Round Trip Times (RTT) since it eliminates performing handshakes repeatedly whenever an endpoint changes IP address. This feature is particularly time-saving when a QUIC endpoint migrates multiple times \cite{liu2024}.

Client-side connection migration is a feature within the IETF QUIC standard that allows clients to change their source IP address while retaining an existing connection with a QUIC server. This process is presented in Fig. \ref{fig:conn_migration}. The CM Initiator (\ie client) detects a source IP address change and sends a \textit{PATH\_CHALLENGE} frame to the CM Responder (\ie server).
The server responds with a \textit{PATH\_RESPONSE} frame, confirming receipt of the challenge and the viability of the new path. The number of probing packet round trips is version-dependent. 
Once the \textit{PATH\_RESPONSE} is received by the client, the client may use the new source IP address for all further (non-probing) communication with the server. Client-side connection migrations are beneficial in cases where a QUIC endpoint moves from Wi-Fi to the cellular network or vice versa \cite{rfc9000}.

\begin{figure}[!b]
\centering
\includegraphics[width=2.1in]{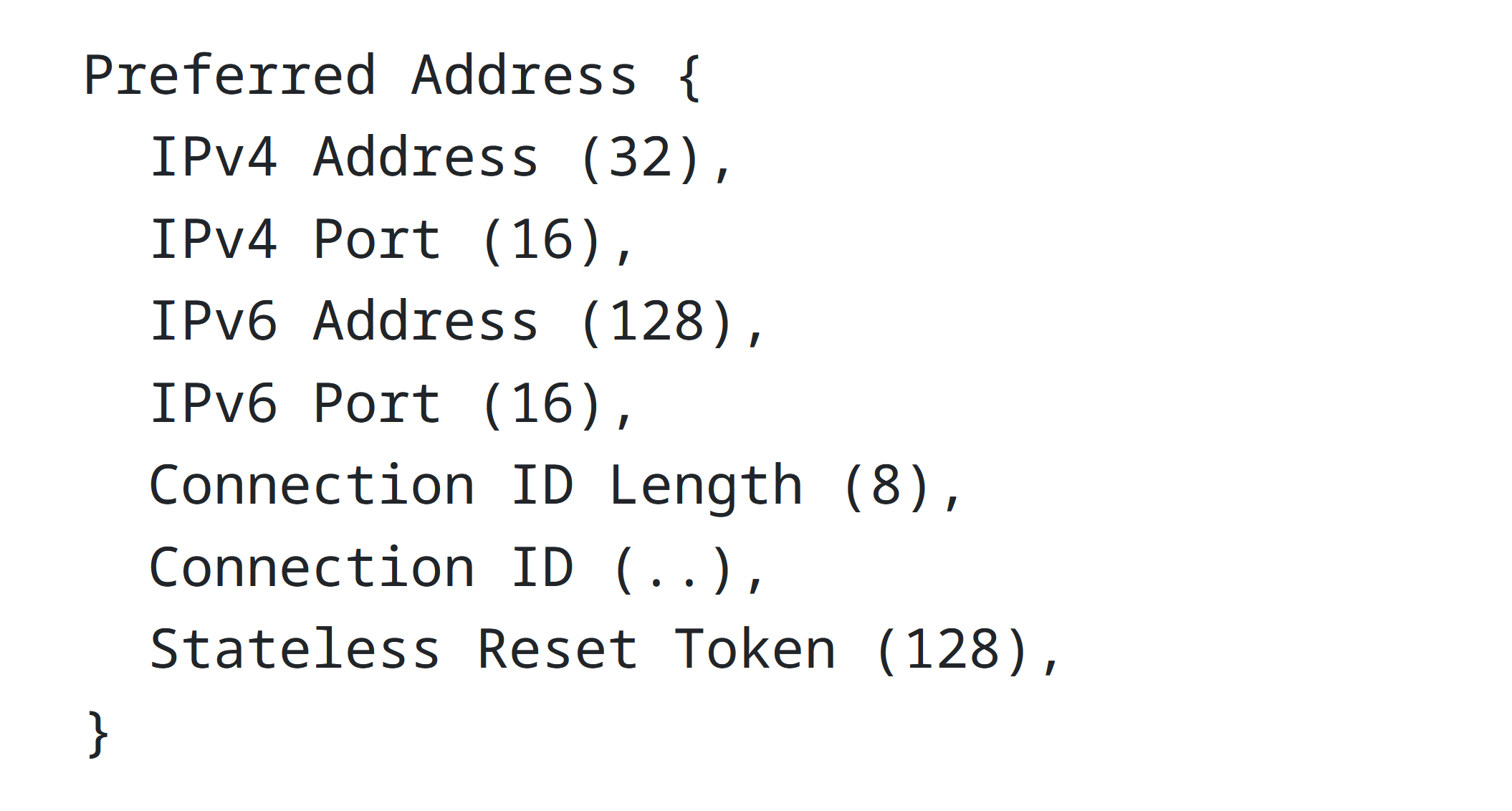}
\caption{Format of the Server Preferred Address Transport Parameter \cite{rfc9000}}
\label{fig:preferred_address}
\end{figure}

Server-side connection migration, like client-side migration, ensures that a QUIC connection can continue seamlessly whenever the server's IP address changes mid-connection. The server \texttt{preferred\_address} transport parameter is a data structure within the \texttt{quic\_transport\_parameters} extension, which defines a secondary (preferred) server address. It is sent from a QUIC server to a client as part of the handshake. The parameter contains fields for an IPv4 Address, IPv4 Port, IPv6 Address, IPv6 Port, CID Length, CID, and the Stateless Reset Token (\cf Fig. \ref{fig:preferred_address}). A server may specify a preferred IPv4 and/or IPv6 address during the initial handshake, that the client can use to communicate with the server at any time after a successful connection establishment to the primary server address. To initiate a server-side connection migration, the client simply sends a path validation packet, containing a \textit{PATH\_CHALLENGE} frame to the secondary address and waits for the server's acknowledgment \cite{rfc9000}. It is important to note that both client-side and server-side connection migrations are initiated by the client. The server-side connection migration can, for example, be used in microservice deployment at the network edge to seamlessly migrate a container hosting a QUIC-based service from one server address to a different one with minimum latency \cite{puliafito2022}. Another use case for server-side connection migration is splitting network traffic mid-session across different network paths to increase privacy. Wang et al.~\cite{wang2022} introduce Connection Migration Powered Splitting (CoMPS), which helps to reduce the risk of traffic analysis attacks by network-level adversaries.

RFC 9000 \cite{rfc9000} recommends reducing the linkability of QUIC connections by, inter alia, using a new CID and source port when migrating to another IP address. Consequently, a middlebox is unable to tell whether a client performed a connection migration, or if a second client is starting to communicate \cite{govil2020}.

The prevalence of QUIC connection migrations has been shown by \cite{buchet2024}, who found that the top providers that support connection migrations are Cloudflare, AWS, Hostinger, Akamai and Google. However, connection migration is not yet supported by other major providers.
\section{Problem Statement}\label{sec:problem_statement}

In this section, we describe how server-side connection migrations can be mimicked to exfiltrate sensitive data from a device. We compare this approach to existing client-side request forgery attacks and highlight the shortcomings of current firewall technologies in analyzing QUIC traffic. 

\subsection{Connection Migration}

As specified in RFC 9000 \cite{rfc9000}, ``\textit{the use of a connection ID allows connections to survive changes to endpoint addresses (IP address and port), such as those caused by an endpoint migrating to a new network.}'' Consequently, both a client and a server can use different IP addresses to communicate while retaining an existing connection \cite{liu2024}.

Further, a ``\textit{server might receive a packet addressed to its preferred IP address at any time after it accepts a connection}'' \cite{rfc9000}. Concretely, a client can choose to migrate the connection to a preferred server destination IP address and initiate the path migration by sending a QUIC packet encapsulating a \textit{PATH\_CHALLENGE} frame to the new (preferred) server IP address. This feature can be exploited by an adversary who wants to mimic a QUIC server-side connection migration by changing the destination IP address of an outgoing QUIC packet and thus sending an illegitimate packet to the adversary's server. The payload of such a packet can be arbitrarily modified so that it encapsulates sensitive data from the victim's machine. As specified in \cite{rfc9000}, the preferred server address is sent from the server to the client as part of the encrypted handshake. Hence, it remains unknown to a middlebox. Therefore, a middlebox is unable to differentiate between legitimate preferred server addresses and malicious ones. It is important to note that migrating to a new path does not require a dedicated handshake \cite{liu2024, shi2019}. The path validation process is solely initiated by the client sending a ``\textit{PATH\_CHALLENGE frame containing an unpredictable payload on the path to be validated}'' \cite{rfc9000}.

Similarly to \cite{gbur2023}, who evaluated the effectiveness of \textbf{client-side} Connection Migration Request Forgery (CMRF) attacks, our method shows that the \textbf{server-side} connection migration feature can be exploited to exfiltrate data from an infected client to a target server.
RFC 9000 \cite{rfc9000} describes ``Request Forgery with Spoofed Migration'' (\ie CMRF) attacks, where clients can spoof the source address of a QUIC packet as part of an apparent connection migration. This tricks the server into sending datagrams to the spoofed address \cite{rfc9000}. However, \cite{rfc9000} fails to mention that spoofing the destination address using the \texttt{preferred\_address} transport parameter of a QUIC packet can be misused to covertly exfiltrate data from a QUIC endpoint to a malicious server --- disguised as a server-side connection migration.

As per RFC 9000 \cite{rfc9000}, the \texttt{preferred\_address} transport parameter may store a single secondary address for each address family (IPv4 and IPv6). There are Internet draft specifications that aim to increase the number of additional addresses that a server can use, such as the Multipath Extension for QUIC by \cite{liu2024}.
According to \cite{piraux2024}, a QUIC frame can be specified which securely advertises additional server IP addresses that a client may use to communicate with the server. Such additional addresses are transmitted inside an \texttt{ADDITIONAL\_ADDRESSES} frame, which is also encrypted with the rest of the handshake and thus remains invisible to a middlebox. Therefore, a malicious exfiltration program may specify an exfiltration server IP address as the destination IP, and a middlebox has to assume that this is a previously advertised server IP address.

\subsection{Firewalls}

It is inherent to the QUIC protocol that CID renegotiation happens in an encrypted way --- out of sight of potential path-based network filtering devices such as firewalls.
In order to keep track of a QUIC connection, a stateful firewall would need to store the Source Connection ID (SCID) during the handshake \cite{gbur2021} and subsequently match the DCID of an outgoing packet with the previously stored SCID. However, changes in DCID are not visible to a middlebox, as renegotiation of CIDs can occur post-handshake inside the encrypted ``Protected Payload'' packets. QUIC does not expose any more unencrypted information on the connection migration process. Since path validation packets are not preceded by a dedicated handshake and the connection can be migrated to another endpoint on both the client and the server-side at any time \cite{rfc9000}, firewalls would need to treat QUIC packets without a prior handshake as connection migration attempts, yielding UDP 5-tuple checks and CID tracking unfeasible.

If no full decryption of the QUIC traffic is done, it is almost impossible for a middlebox to differentiate between a legitimate connection migration and a spoofed one by only analyzing unencrypted packet headers. Currently, firewalls simply filter QUIC traffic based on the handshake packets, but may not be able to perform further stateful inspection after the handshake. 
Therefore, to further motivate this research, we have conducted unstructured interviews with five leading firewall vendors, who support our findings regarding the challenges of effectively analyzing QUIC traffic. They emphasize the difficulty in distinguishing between legitimate and spoofed connection migrations without decrypting QUIC traffic.

In summary, connection migration --- a feature that enhances QUIC's performance --- can also be misused by the data exfiltration method proposed in this paper. In the following section, we present a methodology and the PoC implementation for demonstrating how QUIC's connection migration feature can be exploited and we show that QUIC-based data exfiltration attacks are difficult to differentiate from the normal protocol behavior.
\section{Methodology}\label{sec:methodology}

In the following, we provide a comprehensive description of the proposed data exfiltration attack. After presenting an overview of the threat model and the underlying assumptions, we introduce the methodology, which comprises a sniffing phase, spoofed path validation, and a continued exfiltration phase.

\subsection{Threat Model \& Assumptions}

We consider a scenario in which an attacker has already infected a victim's machine and aims to covertly exfiltrate sensitive data to a target server owned by the attacker. The attack is performed as per the MITRE ATT\&CK framework sub-technique ``Exfiltration Over Alternative Protocol: Exfiltration Over Symmetric Encrypted Non-C2 Protocol''~\cite{mitre}. Circumventing on-device anomaly detection systems is out of the scope of this paper, as we only consider on-path middleboxes and additional on-path traffic analysis software to be the main defenders.

\noindent
The underlying assumptions can be summarized as follows:
\begin{enumerate}
    \item It is assumed that the adversary has successfully gained access to a compromised machine within the internal network of a victim. The compromised machine is \textit{firewalled}, meaning it is located behind a host-based and/or enterprise-level network-based stateful firewall.
    \item The adversary is also capable of secretly deploying the data exfiltration software onto the compromised machine and has gained elevated privileges to successfully execute the attack.
    \item We assume that a defender (\eg the firewall administrator) does not enforce a strict packet inspection policy (\ie full decryption of traffic) and does not outright block UDP ports 80 and 443.
    \item We further assume that the defender knows how to perform (encrypted) traffic analysis, which encompasses the analysis of packet lengths, time deltas between adjacent packets, payload entropy, and the analysis of all the cleartext information sent during the handshake.
\end{enumerate}

\subsection{Data Exfiltration Method}

We base our method on the version-independent properties of the IETF QUIC protocol \cite{rfc8999} and the connection migration feature described in RFC 9000 \cite{rfc9000}, in which a QUIC client has prior knowledge of the alternative server addresses to which it can choose to migrate its connections. During connection establishment, the server notifies the client about its alternative addresses by populating the \texttt{preferred\_address} transport parameter with one or more alternative addresses. If a packet gets lost during the communication, the client can assume that the server has migrated to a new address and the client can choose to validate the new network path by sending a probing packet to the new server address \cite{puliafito2022}. Alternatively, the client may initiate a server-side connection migration at any time by sending a path validation packet to the preferred server address.

\begin{figure}[b]
  \centering
  \includegraphics[width=\linewidth]{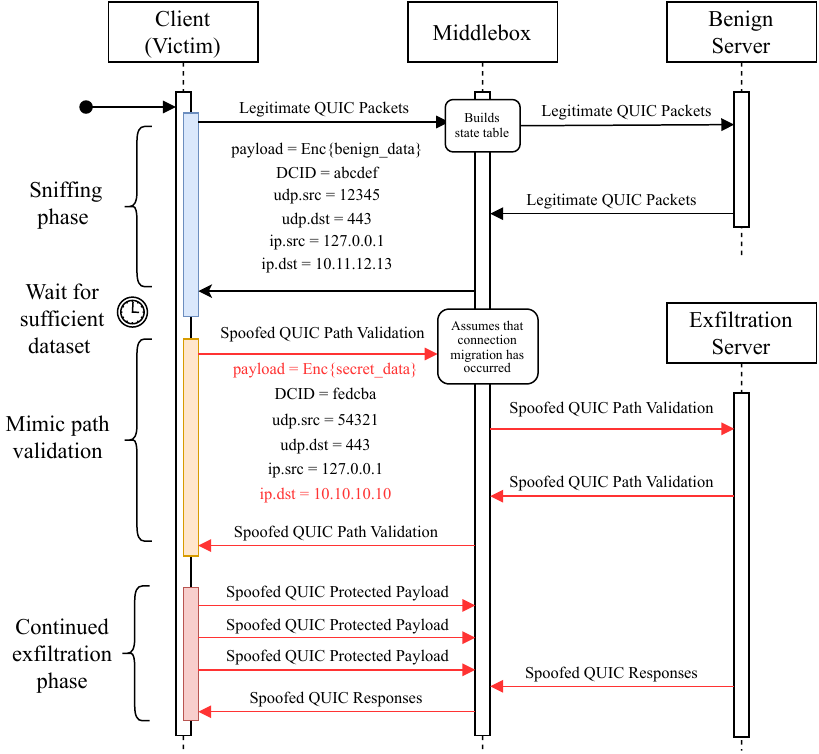}
  \caption{Proposed Data Exfiltration Attack.}
  \label{fig:method}
\end{figure}

Fig. \ref{fig:method} presents the data exfiltration methodology, which is divided into three distinct phases --- a sniffing phase, a path validation phase, and the main exfiltration phase. The involved entities are \1 the client (victim), whose machine is infected with the exfiltration tool, \2 a middlebox, whose task is to detect and block suspicious network traffic, \3 a benign server, and \4 the malicious exfiltration server, which accepts all incoming connections and sends the respective response messages. During the initial sniffing phase, the data exfiltration software collects all outgoing QUIC short header packets and stores relevant information such as IP, UDP, and QUIC headers, as well as timestamps and the QUIC payload lengths. This information is reused at a later stage to mimic the structure of benign packets.

As part of the sniffing phase, the exfiltration tool waits for existing connections to retire their CID by continuously probing whether the socket that binds the currently used source port can be re-bound. Once the CID has been retired, the exfiltration tool can immediately emulate a server-side connection migration, without having to fear that benign traffic continues to be sent and causing suspicious overlaps of benign and malicious traffic originating from the same QUIC connection.

The first step of a connection migration is the path validation, which is mimicked by sending a spoofed path validation request to the malicious exfiltration server. The spoofed path validation packet may already contain the first bytes of sensitive data since a benign path validation packet is also sent via an encrypted channel and is, therefore, not visible to a middlebox. As per \cite{rfc9000}, a benign path validation request includes an 8-byte \textit{PATH\_CHALLENGE} frame, which in turn expects an 8-byte \textit{PATH\_RESPONSE} frame. An endpoint must use datagram sizes of at least 1200 bytes to transmit \textit{PATH\_CHALLENGE} and \textit{PATH\_RESPONSE} frames, as this ensures that the path is able to handle datagrams of this size \cite{rfc9000}. Therefore, the spoofed path validation datagrams are also designed to have a size of at least 1200 bytes. When the middlebox first encounters the path validation packet, it must either assume that a QUIC connection migration is occurring or that an arbitrary UDP packet is sent to a destination address that has not been seen before. In both cases, middleboxes generally allow connections initiated from the inside (trusted) network to the outside.

Lastly, the main exfiltration phase begins by continuously sending packets to the spoofed exfiltration server address and receiving the appropriate spoofed responses from the exfiltration server, to mimic a ``healthy'' connection between a QUIC client and a QUIC server. Thereafter, the malicious connection may then be retired at any time and a new suitable connection can be selected to mimic another connection migration.

It is important to note that the method does not attempt to transmit any handshake packets. Therefore, middleboxes and/or fingerprinting software cannot detect nor block an illegitimate connection establishment attempt based on a handshake packet (\cf Section \ref{subsec:fingerprinting}).

\begin{figure}[t]
  \centering
  \includegraphics[width=0.68\linewidth]{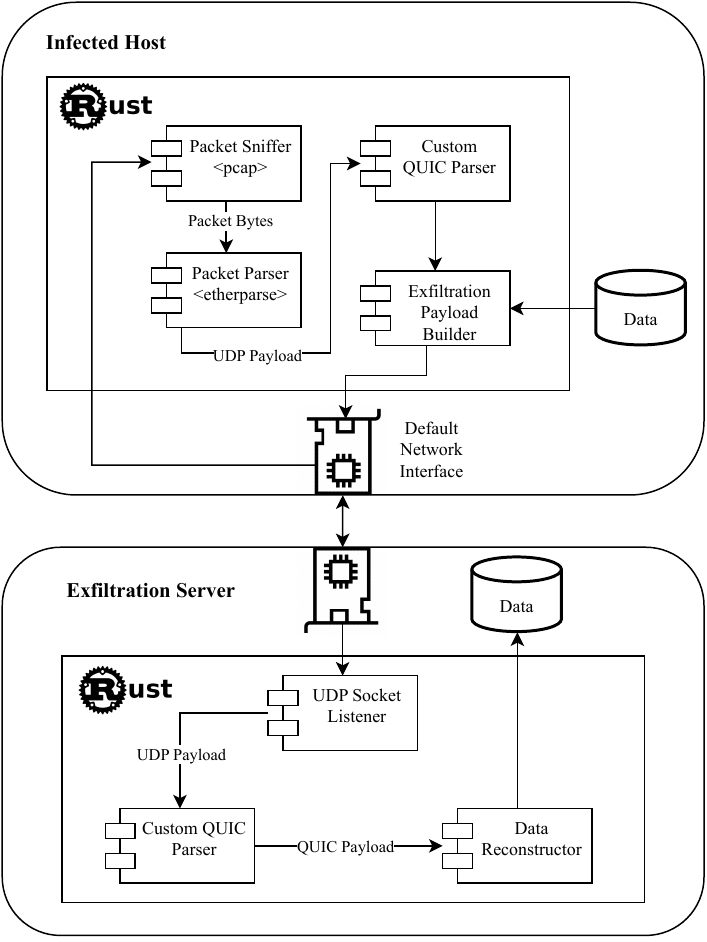}
  \caption{Proof-of-Concept Prototype Architecture.}
  \label{fig:architecture}
\end{figure}

\subsection{Increasing Stealthiness}\label{subsec:stealthiness}

In order to increase stealthiness, the data exfiltration tool can store a range of exfiltration server destination IP addresses; it may only use an address once to mimic a single connection migration and discards the address afterwards.

Furthermore, to minimize the payload size variance between benign and malicious traffic, the malicious payload lengths are sampled from a distribution of benign payload lengths, making a statistical analysis of packet lengths less feasible. Similarly, the time deltas between QUIC packets, which can be considered a unique fingerprint of the application, are also mimicked (\cf Section \ref{subsec:anomaly_detection}). The exfiltration payloads are encrypted with an AES-256 key stored in the data exfiltration executable. This key is not used to establish secrecy, but rather to increase the entropy of the data exfiltration payloads, in order to resemble the entropy of benign QUIC protected payload packets.

Data exfiltration is only attempted when a legitimate connection is retired, \ie when its DCID changes, or stops being used by the QUIC endpoint. Additionally, our method's exfiltration throughput is dependent on the user's activity, and only exfiltrates data while the user is actively browsing the web.

\subsection{Domain Registration Records}

Detecting exfiltration traffic based on mismatches between domain registration records of old and new destination IP addresses is possible. However, as shown in \cite{lu2021}, the WHOIS records may not be publicly accessible under certain jurisdictions. The EU General Data Protection Regulation (GDPR) requires certain WHOIS data to be redacted, which could make this attack more difficult to detect.

Furthermore, an attacker could host an exfiltration server on Amazon Web Services (AWS), resulting in domain registration records in which the organization's name appears identical to those of legitimate Amazon.com traffic. Similarly, YouTube traffic IP addresses have WHOIS records pointing to Google LLC. Data exfiltration servers hosted on Google Cloud would have similar WHOIS records, making it challenging to identify them as malicious based only on WHOIS records.

Checking published IP address ranges specific to a cloud provider, which can be obtained directly from a cloud provider's documentation, is one way to identify exfiltration attempts. For instance, when a QUIC connection to \textit{google.com} migrates to a new destination server hosted in a Google Cloud, with identical domain registration records, it may be possible to cross-check against public IP address ranges of Google Cloud services to identify spoofed connection migration attempts.
However, as of now, there is no indicator that leading firewall vendors offer such functionality in their products to identify connection migrations and differentiate between benign and malicious ones (\cf Section \ref{subsec:survey}).

\section{Implementation}\label{sec:implementation}

The PoC prototype was implemented in Rust as depicted in Fig. \ref{fig:architecture} and the client's source code has been released\footnote{\url{https://github.com/thomasgruebl/quic-exfil}}. It consists of a client-side data exfiltration executable, which comprises a packet sniffer, packet parser, custom QUIC parser, and exfiltration payload builder module. The server-side executable consists of a UDP socket listener, a custom QUIC parser and a data reconstruction module.

Every exfiltration task is performed in a new thread, meaning that multiple malicious connection migrations can be performed in parallel for different QUIC connections, while the outgoing benign traffic is continuously monitored for suitable connections that can be used to mimic new connection migrations.
The outgoing QUIC packets are sniffed using the Rust \textit{pcap} library \cite{pcap} (version 1.2.0) and parsed using the Rust \textit{etherparse} library \cite{etherparse} (version 0.14.2). It is assumed that QUIC uses port 443 over UDP. Every outgoing UDP packet's payload is parsed as per the version-independent properties of the QUIC protocol \cite{rfc8999}. The header structure of short header packets does not specify where the DCID ends nor does it specify the length of the DCID. Hence, observing a short header packet alone, does not give any indication to where the DCID bytes end and where the payload bytes start. 
If the parser identifies a QUIC long header packet as part of a handshake, it extracts its DCID, stores it, and subsequently discards the packet. The DCID is then used to map succeeding short header packets to their preceding handshake. This is required to \1 replicate the exact length of the original DCID in spoofed packets, as QUIC endpoints usually stick to a certain CID length, and \2 mimic the lengths of the original payloads. Special cases, such as packets that contain a long and a short header at the same time or packets that contain a zero-length DCID, are discarded.

Monitoring all outgoing packets on the default network interface can result in a loop, as the exfiltration client itself also transmits packets over this interface. Sniffing and parsing those packets would result in an unnecessary performance overhead and would dilute the observed features of benign packets. To prevent this, the exfiltration client maintains a blacklist to ignore all malicious packets. This blacklist contains the SHA-3-256 hashes of all previously sent malicious packets, against which every outgoing packet is checked.

\section{Evaluations}\label{sec:evaluation}

\begin{table*}[h]
\centering
\caption{Network Traffic Generation and Device Roles in the Experimental Testbed.}
\scalebox{0.9} {
\begin{tabular}{P{2.5cm} | P{0.5cm} P{0.5cm} P{0.5cm} P{0.5cm} P{0.5cm} P{0.5cm} P{0.5cm} P{0.5cm} P{0.5cm} P{0.5cm} P{0.5cm} P{0.5cm} P{0.5cm} P{0.5cm} P{0.5cm} P{0.5cm} }
\toprule
\multicolumn{16}{c}{\textbf{\textit{Device Number}}} \\ \midrule
 \textbf{\textit{Traffic Type}} & \textbf{1} & \textbf{2} & \textbf{3} & \textbf{4} & \textbf{5} & \textbf{6} & \textbf{7} & \textbf{8} & \textbf{9} & \textbf{10} & \textbf{11} & \textbf{12} & \textbf{13} & \textbf{14} & \textbf{15} & \textbf{16} \\ \midrule

QUIC Traffic & \tikz \fill (0,0) circle [radius=0.1cm]; & \tikz \fill (0,0) circle [radius=0.1cm]; & \tikz \fill (0,0) circle [radius=0.1cm]; & \tikz \fill (0,0) circle [radius=0.1cm]; & \tikz \fill (0,0) circle [radius=0.1cm]; & \tikz \fill (0,0) circle [radius=0.1cm]; & \tikz \fill (0,0) circle [radius=0.1cm]; & \tikz \fill (0,0) circle [radius=0.1cm]; & \tikz \draw (0,0) circle [radius=0.1cm]; & \tikz \draw (0,0) circle [radius=0.1cm]; & \tikz \draw (0,0) circle [radius=0.1cm]; & \tikz \draw (0,0) circle [radius=0.1cm]; & \tikz \draw (0,0) circle [radius=0.1cm]; & \tikz \draw (0,0) circle [radius=0.1cm]; & \tikz \draw (0,0) circle [radius=0.1cm]; & \tikz \draw (0,0) circle [radius=0.1cm]; \\ \midrule

Benign Migration & \tikz \fill (0,0) circle [radius=0.1cm]; & \tikz \fill (0,0) circle [radius=0.1cm]; & \tikz \fill (0,0) circle [radius=0.1cm]; & \tikz \fill (0,0) circle [radius=0.1cm]; & \tikz \fill (0,0) circle [radius=0.1cm]; & \tikz \fill (0,0) circle [radius=0.1cm]; & \tikz \fill (0,0) circle [radius=0.1cm]; & \tikz \fill (0,0) circle [radius=0.1cm]; & \tikz \draw (0,0) circle [radius=0.1cm]; & \tikz \draw (0,0) circle [radius=0.1cm]; & \tikz \draw (0,0) circle [radius=0.1cm]; & \tikz \draw (0,0) circle [radius=0.1cm]; & \tikz \draw (0,0) circle [radius=0.1cm]; & \tikz \draw (0,0) circle [radius=0.1cm]; & \tikz \draw (0,0) circle [radius=0.1cm]; & \tikz \draw (0,0) circle [radius=0.1cm]; \\ \midrule

Exfiltration & \tikz \draw (0,0) circle [radius=0.1cm]; & \tikz \draw (0,0) circle [radius=0.1cm]; & \tikz \draw (0,0) circle [radius=0.1cm]; & \tikz \draw (0,0) circle [radius=0.1cm]; & \tikz \draw (0,0) circle [radius=0.1cm]; & \tikz \draw (0,0) circle [radius=0.1cm]; & \tikz \fill (0,0) circle [radius=0.1cm]; & \tikz \fill (0,0) circle [radius=0.1cm]; & \tikz \fill (0,0) circle [radius=0.1cm]; & \tikz \fill (0,0) circle [radius=0.1cm]; & \tikz \draw (0,0) circle [radius=0.1cm]; & \tikz \draw (0,0) circle [radius=0.1cm]; & \tikz \draw (0,0) circle [radius=0.1cm]; & \tikz \draw (0,0) circle [radius=0.1cm]; & \tikz \draw (0,0) circle [radius=0.1cm]; & \tikz \draw (0,0) circle [radius=0.1cm]; \\ \midrule

Non-QUIC Traffic & \tikz \fill (0,0) circle [radius=0.1cm]; & \tikz \fill (0,0) circle [radius=0.1cm]; & \tikz \fill (0,0) circle [radius=0.1cm]; & \tikz \fill (0,0) circle [radius=0.1cm]; & \tikz \fill (0,0) circle [radius=0.1cm]; & \tikz \fill (0,0) circle [radius=0.1cm]; & \tikz \fill (0,0) circle [radius=0.1cm]; & \tikz \fill (0,0) circle [radius=0.1cm]; & \tikz \fill (0,0) circle [radius=0.1cm]; & \tikz \fill (0,0) circle [radius=0.1cm]; & \tikz \fill (0,0) circle [radius=0.1cm]; & \tikz \fill (0,0) circle [radius=0.1cm]; & \tikz \fill (0,0) circle [radius=0.1cm]; & \tikz \fill (0,0) circle [radius=0.1cm]; & \tikz \fill (0,0) circle [radius=0.1cm]; & \tikz \fill (0,0) circle [radius=0.1cm]; \\ \midrule

\end{tabular}
}
\label{tab:experimental_setup}
\end{table*}

This section first introduces the experimental setup and the three different user activity scenarios that were considered. Subsequently, this section focuses on identifying unique features of QUIC connection migrations and training anomaly detection classifiers to try to detect the proposed attack. The remainder of this section discusses the feasibility of detecting the proposed attack using fingerprinting tools, followed by the results of a survey of leading firewall vendors, who were asked to assess the QUIC-filtering capabilities of their firewall products as part of unstructured interviews.

\subsection{Experimental Setup}

In our experimental setup, we simulate a small network of devices to generate QUIC traffic, including benign and malicious server-side connection migrations. The network consists of 16 Docker containers running Ubuntu 18.04 LTS with the Xfce desktop environment and VNC/noVNC servers to provide remote access \cite{dockercontainer}. Our machine acting as a firewall is a virtual machine (VM) with 32 GB of RAM, running Ubuntu 22.04 LTS. It uses iptables with the conntrack module for stateful packet filtering and connection tracking.
Packet capturing is performed on the inward-facing interface of the firewall (\ie the Docker virtual Ethernet bridge adapter). As shown in Table \ref{tab:experimental_setup}, the network generates multiple types of traffic. Devices 1--8 generate QUIC traffic, including benign QUIC connection migrations to simulate real-world user-initiated network traffic.  Devices 9--16 represent other backend (non user-controlled) machines, which do not generate QUIC traffic. Devices 7--10 are infected with the data exfiltration software and generate malicious connection migrations as well as QUIC-based data exfiltration traffic. All devices (1--16) also generate other types of non-QUIC network traffic. To emulate QUIC server-side connection migrations, we used a modified version of Cloudflare \textit{quiche} v0.23.2 \cite{cloudflare-quiche}. The quiche server is running on the firewall, the quiche clients are running on devices 1--8 and trigger QUIC connection migrations at random time intervals every 0--30 minutes. This allows us to collect benign connection migration fingerprints. The total number of generated connection migrations, benign and malicious, are presented in Table \ref{tab:connection_migrations}. Every single malicious connection migration in the dataset, that is followed by at least one exfiltration payload packet, constitutes a data exfiltration attempt.

\begin{table}[t]
\centering
\caption{Number of Benign and Malicious Connection Migrations per Scenario.}
\scalebox{0.9} {
\begin{tabular}{>{\centering\arraybackslash}m{1.8cm} | *{3}{>{\centering\arraybackslash}m{1.8cm}}} 
\toprule
\textbf{\textit{Scenario}} & \textbf{\textit{Benign}} & \textbf{\textit{Malicious}} \\ 
\midrule
Mixed & 245 & 27 \\ \midrule
YouTube & 371 & 34 \\ \midrule
Noise & 98 & 11 \\ \midrule
\midrule
Total & 714 & 72 \\
\bottomrule
\end{tabular}
}
\medskip
\label{tab:connection_migrations}
\end{table}

As mentioned in Section \ref{subsec:stealthiness}, to increase the stealthiness of the method, the exfiltration throughput depends on the network activity of the infected hosts. Therefore, the following evaluation does not demonstrate that a certain throughput can be achieved. Instead, it considers three different user scenarios in which the exfiltration tool attempts to send data to a target server and evaluates the performance of the anomaly detection classifiers. The following user activity scenarios were considered:

\begin{enumerate}
    \item General Network Activity (24h of Mixed Traffic): A simulated small company network consisting of 16 Docker containers, each representing a host that collectively generates web traffic over a 24-hour period. This includes browsing across various domains (\eg youtube.com, google.com, facebook.com, instagram.com, cloudflare.com, amazon.com, chatgpt.com) and thus creating a number of different overlapping QUIC connections.
    \item Isolated Streaming Scenario (24h of YouTube Traffic): The same simulated network, but now restricted to generating primarily YouTube video traffic over a 24-hour period. This scenario serves as a controlled test case to contrast with the more diverse traffic patterns of the general network activity.
    \item Background Noise Traffic (24h Idle Mode): A subset of devices generating low-interaction background traffic for 24 hours, simulating idle devices with existing open QUIC connections (\eg open browser tabs, apps, etc.) but minimal active user interaction.
\end{enumerate}

Across all three scenarios, we gathered a total of 710,690 outgoing QUIC short header packets, of which 690,101 are benign and 20,589 are spoofed. Scenario 1 (\ie mixed traffic) generated 427,644 outgoing QUIC short header packets, 416,961 of which are benign and 10,683 of which are spoofed. The total achieved exfiltration volume is 6.34 MB. Scenario 2 (\ie YouTube traffic) generated 255,649 outgoing QUIC short header packets, 247,624 of which are benign and 8,025 of which are spoofed, resulting in a total data exfiltration volume of 2.97 MB. Scenario 3 (\ie noise traffic) generated a total of 27,397 packets, including 25,516 benign and 1,881 malicious ones and achieved a total volume of 1.10 MB.

\begin{table*}[t]
    \centering
    \caption{Anomaly Detection Features.}
    \label{tab:features}
    \scalebox{0.75} {
    \begin{tabular}{@{}c|cc p{12cm} @{}}
        \toprule
        \textbf{\textit{Category}} & \textbf{\textit{Feature}} & \textbf{\textit{Considered}} & \textbf{\textit{Justification}}\\
         \midrule
        \multirow{2}{*}{Handshake}& TLS Client Hello, Server Hello, etc. & \xmark & Both legitimate and spoofed connection migrations cannot be reliably mapped to a preceding handshake.\\
         \midrule
         \multirow{18}{*}{Short Header Packets}& Packet Length & \xmark & The length of the entire packet is dependent on features that may also differ across benign packets (\eg IP header structure). \\
        & SCID/DCID & \xmark & Both the length of the CIDs and the CIDs themselves have no informational content and can vary across benign and malicious traffic.\\
        & QUIC Payload Length & \checkmark & The payload length may be used as an indicator of irregular traffic patterns. Unusual payload sizes can indicate data exfiltration attempts.\\
        & Connection Migration Payload Length & \checkmark & The payload length of a connection migration packet (\ie a packet containing a \textit{PATH\_CHALLENGE} frame) can be used as an indicator of unusual connection migration attempts.\\
        & Time $\Delta$ between two outgoing packets & \checkmark & The data exfiltration tool may have different processing times compared to a benign QUIC endpoint.\\
        & Time $\Delta$ between two ingoing packets & \xmark & This feature is similar to the previous one, since the exfiltration server's QUIC endpoint would be functioning identically to the QUIC endpoint running on the client, therefore only one of the two features need to be considered.\\
        & Time $\Delta$ between a request-response pair & \xmark & The RTT latency can be arbitrary for both legitimate and spoofed addresses, depending on the physical location of the destination server. \\
        & Payload Entropy & \xmark & The entropy of payloads can be imitated by encrypting the exfiltration payload using a cryptographic key stored in the exfiltration software.\\
        & Latency Spin Bit & \xmark & The latency spin bit is used by on-path observers to measure the per-round-trip latency. For every received packet from the server, the client simply flips the bit, which does not provide enough informational content to differentiate benign from malicious traffic.  \\
        & Fixed Bit & \xmark & As the name suggests, the fixed bit has a constant value and serves as a way to differentiate QUIC traffic from other UDP-based traffic. It can simply be adopted by malicious packets.  \\
        & Packet number & \xmark & In short header packets, the packet number, as well as its length, are cryptographically obfuscated and thus not visible to middleboxes.  \\
        & Other QUIC-version dependent bits & \xmark & A spoofed packet can simply adopt these bits from a benign packet. \\
        \bottomrule
    \end{tabular}
    }
\end{table*}

It is evident that the data exfiltration throughput is highly dependent on the victim's activity. In cases where the user browses websites that do not require large amounts of data to be downloaded, only infrequent QUIC acknowledgment packets are being sent back to the server. YouTube traffic generates many outgoing QUIC acknowledgments with small payload sizes. Data uploads to a cloud provider typically generate a high number of outgoing QUIC packets using the maximum payload size. Mimicking a connection migration in the latter case enables the attacker to exfiltrate large amounts of sensitive data.

\subsection{Anomaly Detection}\label{subsec:anomaly_detection}

In this evaluation section, we analyze the proposed data exfiltration method by assessing the ability of machine learning-based anomaly detectors to detect this kind of malicious network behaviour. Since existing anomaly detection classifiers that detect network-based exfiltration attempts are commonly trained on handshake metadata \cite{zhan2022}, and QUIC connection migrations do not require handshakes, a new feature set needs to be defined for QUIC-based data exfiltration traffic.

Table \ref{tab:features} presents a range of different features that can be considered in the context of QUIC traffic analysis. Most features can be easily imitated --- only three features have been identified which require some level of sophistication to imitate. Therefore, our objective was to mimic the following features with the help of our PoC implementation: \circled{1} The connection migration payload, \circled{2} the ``normal'' QUIC packet payload length, and \circled{3} the time delta between two outgoing packets of the same QUIC flow.

Feature \circled{1} is a binary feature that captures connection migration attempts in outgoing QUIC packets. More specifically, it observes the first packet of a connection migration and stores its payload length. We classify every observed migration attempt accordingly, using this binary label as a potential discriminator between ``normal'' protected payload packets and packets containing a \textit{PATH\_CHALLENGE} frame.

Feature \circled{2}, payload length, can be imitated by first observing benign traffic during the sniffing phase depicted in Fig. \ref{fig:method}. After a sufficiently sized dataset of benign QUIC payload lengths has been collected, the data exfiltration tool randomly samples from this dataset -- essentially replicating the payload sizes of previously seen traffic of the same QUIC flows. For instance, when mimicking a connection migration of a YouTube server, the exfiltration tool continues to use the same payload sizes as those observed in benign acknowledgment packets sent from the client to the YouTube server. We heuristically chose to gather a dataset of 1,000 packets per QUIC connection prior to the first exfiltration attempt.

To replicate the payload sizes of benign QUIC flows, we define $\mathcal{D}$ to be the entire dataset that is created over the course of an exfiltration process. Since the dataset changes over time, we denote the subset of the dataset at time $t$ by $\mathcal{D}_t$ which is used to choose the next QUIC payload size. We then randomly sample a value $x$ from the dataset $\mathcal{D}$ at time $t$,

\begin{center}
$x_t \sim \mathcal{D}_t$
\end{center}

and use this value $x_t$ to set the length of the next malicious payload.

\begin{algorithm}[t]
\caption{Compute Time Deltas per DCID}\label{alg:compute_time_deltas}
\begin{algorithmic}[1]
\Require Pre-filled hashmap $H$ \Comment{Key: DCID, Value: Sorted list of timestamps}

\State Initialize an empty hashmap $\Delta T$ \Comment{Stores time deltas for all DCIDs}

\For{each ($DCID$, $T_{\text{list}}$) in $H$}
    \If{$|T_{\text{list}}| > 1$}
        \For{$i \gets 1$ to $|T_{\text{list}}| - 1$}
            \State Compute time delta:
            \[
            \Delta T_i = T_{\text{list}}[i+1] - T_{\text{list}}[i]
            \]
            \State Append $\Delta T_i$ to $\Delta T[DCID]$
        \EndFor
    \EndIf
\EndFor

\noindent
\State \textbf{return} $\Delta T$

\end{algorithmic}
\end{algorithm}

\begin{algorithm}[t]
\caption{Mimic Observed Time Deltas}\label{alg:mimic_time_deltas}
\begin{algorithmic}[1]
\Require $\Delta T = \{ DCID_1 : [dt_1, dt_2, ...], DCID_2 : [dt_1, dt_2, ...] \}$ \Comment{Observed time deltas}
\Require $BR$ \Comment{Base rate of packet sending}
\Require $n$ \Comment{Number of packets to exfiltrate}
\Ensure Adjusted sleep times for mimicking the observed rate
\For{$i \gets 1$ to $n$}
    \State $\Delta T_i = $ Get random sample from $\Delta T[DCID]$
    \State Compute sleep time:
    \[
    S_i = \Delta T_i - BR
    \]
    \If{$S_i > 0$}
        \State Sleep for $S_i$ milliseconds
    \Else
        \State No sleep needed
    \EndIf
    \State Send packet
\EndFor
\end{algorithmic}
\end{algorithm}

Feature \circled{3}, the time delta between two outgoing packets, is computed as per Algorithm \ref{alg:compute_time_deltas}. The intervals in which the packets are placed ``on the wire'' can be considered a unique fingerprint of an application. From the attacker's perspective, this is the most difficult feature to mimic among the three, because it requires dynamic adjustments of inter-arrival times on a per-flow basis. To mimic feature \circled{3}, Algorithm \ref{alg:mimic_time_deltas} was implemented.

Algorithm \ref{alg:compute_time_deltas} monitors the time deltas of benign QUIC connections and stores them in a dictionary. During data exfiltration, Algorithm \ref{alg:mimic_time_deltas} randomly samples from this distribution to determine the appropriate sleep time before sending the next packet. The base rate of exfiltration was determined by running a single thread of the tool, measuring the time deltas, and averaging them. Since the standard deviation is low, the average can be reliably used. The base rate is typically very close to zero (median: 7 ms, average: 58 ms), meaning the required sleep times closely match the observed time deltas. This low base rate can be traced back to the fact that the exfiltration tool bypasses common rate-limiting mechanisms such as flow control and congestion control, making it faster than most other applications.

We do not consider time intervals between adjacent pairs of requests and responses, because these are dependent on the location of the destination server -- which can also significantly differ across benign servers performing connection migrations. Nor do we consider the time deltas between two ingoing packets, since this feature's expressiveness is identical to \circled{3}.

\begin{table}[t]
\centering
\caption{Comparison of the Classification Performance Across Three Scenarios}
\label{tab:anomaly_detection_results}
\scalebox{0.9}{
\begin{tabular}{>{\centering\arraybackslash}m{1.4cm} | >{\centering\arraybackslash}m{1.5cm} | *{3}{>{\centering\arraybackslash}m{1.5cm}}}
\toprule
\multirow{3}{*}{\textbf{\textit{Classifier}}} & \multirow{3}{*}{\textbf{\textit{Metric}}} & \multicolumn{3}{c}{\textbf{\textit{Scenario}}} \\
\cmidrule(lr){3-5}
& & \textbf{\textit{Mixed}} & \textbf{\textit{Youtube}} & \textbf{\textit{Noise}} \\
\midrule
\multirow{4}{*}{RF}
& F1-Score & 0.35 & 0.18 & 0.47 \\
& Recall & 0.31 & 0.15 & 0.45 \\
& Precision & 0.40 & 0.22 & 0.50 \\
& Accuracy & 0.97 & 0.96 & 0.94 \\
\midrule
\multirow{4}{*}{MLP}
& F1-Score & 0.07 & 0.10 & 0.20 \\
& Recall & 0.44 & 0.34 & 0.45 \\
& Precision & 0.04 & 0.06 & 0.13 \\
& Accuracy & 0.72 & 0.81 & 0.78 \\
\midrule
\multirow{4}{*}{SVM} & F1-Score & 0.02 & 0.05 & 0.08 \\
& Recall & 0.03 & 0.07 & 0.08 \\
& Precision & 0.02 & 0.04 & 0.08 \\
& Accuracy & 0.93 & 0.92 & 0.89 \\
\midrule
\multirow{4}{*}{AE} & F1-Score & 0.00 & 0.01 & 0.00 \\
& Recall & 0.01 & 0.01 & 0.00 \\
& Precision & 0.00 & 0.01 & 0.01 \\
& Accuracy & 0.92 & 0.92 & 0.93 \\
\midrule
\multirow{4}{*}{IF} & F1-Score & 0.06 & 0.08 & 0.05 \\
& Recall & 0.23 & 0.21 & 0.08 \\
& Precision & 0.03 & 0.05 & 0.04 \\
& Accuracy & 0.82 & 0.84 & 0.83 \\
\bottomrule
\end{tabular}
}
\medskip
\footnotesize RF $\rightarrow$ Random Forest, MLP $\rightarrow$ Multi-Layer Perceptron, SVM $\rightarrow$ Support Vector Machine, AE $\rightarrow$ Autoencoder , IF $\rightarrow$ Isolation Forest
\end{table}

\begin{figure*}[h]
    \centering
    \begin{minipage}[b]{0.5\textwidth}
        \centering
        \includegraphics[width=0.7\textwidth]{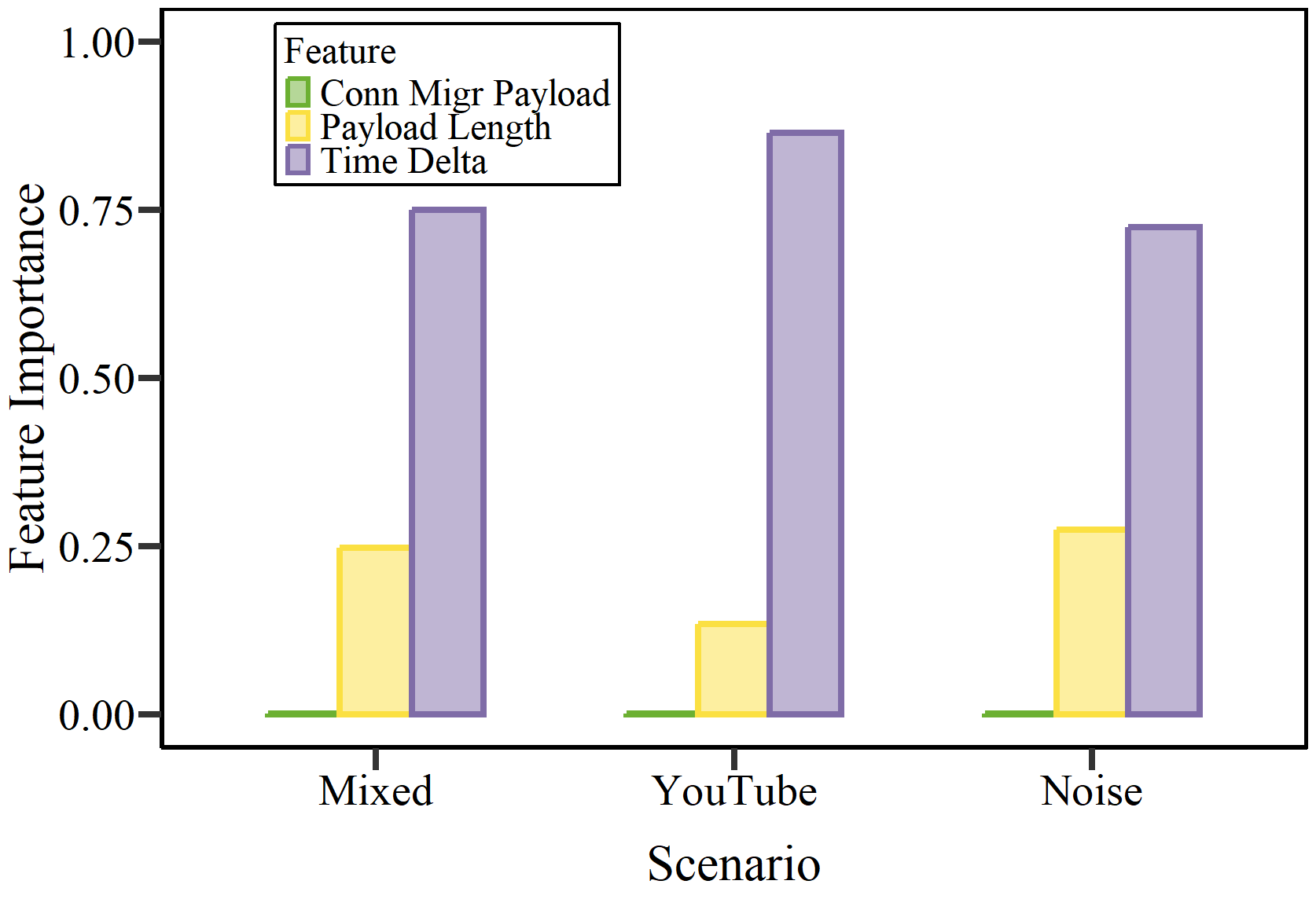}
        \subcaption{Random Forest Classifier.}
        \label{fig:rf_feature_importance}
    \end{minipage}%
    \begin{minipage}[b]{0.5\textwidth}
        \centering
        \includegraphics[width=0.7\textwidth]{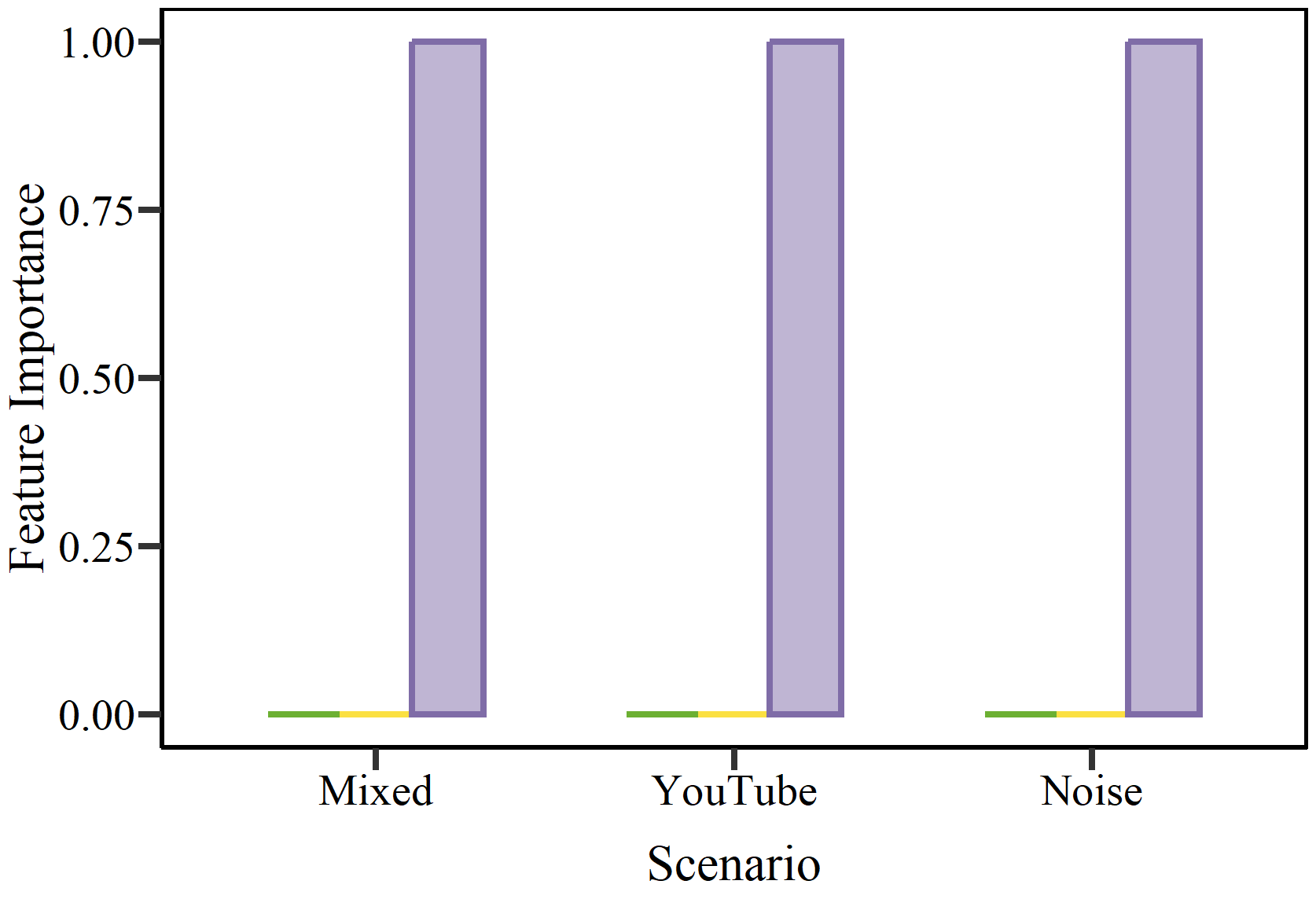}
        \subcaption{Multi-Layer Perceptron Classifier.}
        \label{fig:mlp_feature_importance}
    \end{minipage}
    \caption{Normalized Feature Importance Across Three Scenarios.}
\end{figure*}

Given the inherent challenges of anomaly detection in realistic highly imbalanced network traffic datasets where data exfiltration attempts constitute a small fraction ($\sim$3\%) of the overall traffic, we evaluated the performance of five distinct anomaly detection classifiers. These classifiers use both supervised methods --- Random Forest (RF), Multi-Layer Perceptron (MLP), and Support Vector Machine (SVM) --- and unsupervised methods --- Autoencoder (AE) and Isolation Forest (IF). Supervised methods, such as RF, MLP, and SVM, typically perform well in scenarios where clear decision boundaries can be learned from labeled examples \cite{zhan2022}. Unsupervised methods, like AE and IF, are designed to identify anomalies without explicit labels, by learning the characteristics of normal data and flagging deviations. We selected these classifiers to assess the robustness of feature-based anomaly detection against the data exfiltration technique designed to mimic benign traffic patterns.

We evaluated the detection performance across the three user activity scenarios. Table \ref{tab:anomaly_detection_results} presents a comparative analysis of the classifiers, which focuses on the following metrics: Precision, Recall, and F1-Score for the anomaly class, and overall Accuracy. Due to the class imbalance, accuracy alone is not a reliable indicator of performance, as high accuracy can be trivially achieved by classifiers that predominantly predict the majority class (\ie benign traffic).

Across all scenarios, the results presented in Table \ref{tab:anomaly_detection_results} reveal that all five classifiers are unable to effectively detect the data exfiltration attempts. Despite achieving high accuracy scores in some scenarios, these scores are misleading due to the imbalanced datasets. The more important metrics for the anomaly class --- F1-Score, Recall, and Precision --- consistently show very low values across all classifiers and scenarios.

In the YouTube scenario, for instance, the RF classifier achieves a relatively high accuracy of 96\%. However, its F1-Score for detecting data exfiltration is 0.18, with a Recall of only 0.15 and a Precision of 0.22. This indicates that while RF correctly classifies the majority of benign traffic, it fails to identify most actual data exfiltration attempts and produces a high number of false negatives. Similar poor performance is observed for the MLP and SVM classifiers in the YouTube scenario, with even lower F1-Scores of 0.10 and 0.05 respectively. The AE yields a near-zero F1-Score (0.01) and IF achieves a slightly higher but still very low F1-Score of 0.08.

In the Noise scenario, RF achieves a high accuracy of 96\%, while its F1-Score for anomaly detection remains low at 0.47. MLP performs even worse with an F1-Score of 0.20 in this scenario. SVM and AE again show very poor anomaly detection performance, with near-zero F1-Scores. IF exhibits a slightly improved F1-Score of 0.05 compared to AE and SVM.

In the Mixed traffic scenario, the anomaly detection performance of all classifiers is low. RF achieves the highest F1-Score among the classifiers at 0.35, but still fails to differentiate benign from exfiltration traffic. MLP, SVM, AE, and IF all exhibit very low F1-Scores, ranging from 0.00 to 0.07.

We have also conducted a feature importance analysis for both the RF and MLP detectors. We used the \textit{Gini} importance metric for RF, a measure of how much each feature contributes to the homogeneity of nodes in decision trees. For the MLP detector, we employed the Sharpley Additive Explanation (SHAP) method \cite{NIPS2017_7062}, a game-theoretical approach that attributes the model output to each feature based on their marginal contributions. Fig. \ref{fig:rf_feature_importance} reveals that feature \circled{3} is the most important across all three scenarios. Our analysis also indicates that for the MLP, the time delta feature contributes 100\% to its classification decisions (\cf Fig. \ref{fig:mlp_feature_importance}). However due to the successful mimicking strategy of time delta feature \circled{3}, it is still difficult to detect malicious attempts.

\subsection{Fingerprinting Software}\label{subsec:fingerprinting}

\begin{table*}[t]
\centering
  \caption{Perspectives on the QUIC Protocol from Leading Firewall Vendors}
  \label{tab:survey_firewall_vendors}
  
  \scalebox{0.9}{
  \begin{tabular}{@{}l|P{1cm}P{1cm}P{1cm}P{1cm}P{1cm} @{}}
        \toprule
        \textbf{\textit{Firewall vendor ...}} & \textbf{\textit{A}} & \textbf{\textit{B}} & \textbf{\textit{C}} & \textbf{\textit{D}} & \textbf{\textit{E}}\\
    \midrule
    \1 recommends blocking QUIC entirely.&\checkmark&\checkmark&\checkmark&\checkmark&\xmark\\
    \2 builds a state table based on QUIC connection IDs.&\xmark&\xmark&\xmark&\xmark&\xmark\\
    \3 can differentiate QUIC traffic from other types of traffic.&\checkmark&\checkmark&\checkmark&\checkmark&\checkmark\\
    \4 can perform basic filtering of QUIC traffic (allow / deny).&\checkmark&\checkmark&\checkmark&\checkmark&\checkmark\\
    \5 currently offers functionality to decrypt QUIC traffic on the firewall.&\xmark&\xmark&\checkmark&\checkmark&\checkmark\\
    \6 has HTTP/3 Deep Packet Inspection (DPI) capabilities.&\xmark&\xmark&\checkmark&\checkmark&\checkmark\\
    \7 can recognize QUIC connection migrations.&\xmark&\xmark&\xmark&\xmark&\xmark\\
    \8 plans on / is currently developing new QUIC-related firewall features.&\checkmark&\checkmark&\checkmark&$\sim$&\checkmark\\
    \9 believes that the competitors struggle with identical challenges.&\checkmark&\checkmark&$\sim$&\checkmark&\checkmark\\
\bottomrule
\end{tabular}
}
 \par\medskip
    \footnotesize
     $\sim$~$\rightarrow$ ~Is unsure / cannot provide clear answer.
\end{table*}

Network metadata fingerprinting tools collect information from packet headers (\eg IP addresses, port numbers, TLS-specific metadata) to create a ``fingerprint'' that uniquely identifies a network packet flow. This fingerprint can be used not only for various purposes, such as network performance optimization and device management, but also to make inferences about the underlying application or website that is visited. The features used for fingerprinting can be extracted by a passive eavesdropper, such as an on-path observer \cite{goldberg2020}. Fingerprinting tools, such as the open-source tools Cisco Mercury \cite{mercury} or FATT \cite{fatt}, extract a fingerprint from a QUIC handshake and try to map non-handshake packets to their preceding handshakes. The mapping between the handshakes and the corresponding payload packets is done using the 5-tuple. Packets with modified 5-tuples and no preceding handshakes (\ie QUIC connection migration packets) are therefore not mapped to a specific fingerprint.

Wireshark \cite{wireshark} offers the ability to follow QUIC streams. When encountering unknown QUIC packets that cannot be associated with a handshake, the packets are labeled as ``\textit{Unknown QUIC connection. Missing Initial Packet or migrated connection?}'', if the protected payload packet contains a CID that was not previously seen as part of the handshake packets.
In our current PoC implementations, we reuse existing DCIDs. Even when new DCIDs are used, Wireshark's labeling process cannot distinguish between benign and malicious connection migrations.

As part of the evaluation, a post-analysis of the captured PCAP files has been performed using Cisco Mercury \cite{mercury}. The preceding handshake for both benign and malicious connection migrations was detected by Cisco Mercury, however, the tool fails to map the traffic after a migration (including the migration event itself) to the original handshake. This means, that the tool is unable to correlate the post-migration traffic with the initial connection, hence, treating it as an entirely new flow rather than a continuation of the original session. As a result, since the new UDP flow lacks handshake data, the malicious destination IP address of the exfiltration server cannot be found in the Cisco Mercury fingerprinting results.

\subsection{Survey of Leading Firewall Vendors}\label{subsec:survey}

As part of this study, unstructured interviews with leading firewall vendors were conducted. The qualitative approach of surveying vendors was intended to replace a quantitative evaluation because it directly addresses the core question: whether modern firewalls even possess the necessary features to reliably detect QUIC connection migrations. We contacted eight firewall vendors listed in \cite{gartner}, five of which agreed to share their perspectives on the QUIC protocol for research purposes. The interviewees' roles included ``Systems Engineer'', ``Cyber Security Specialist'', ``Senior Sales Engineer'', ``Consulting Systems Engineer'', ``Senior Solutions Engineer'' and ``Systems Engineering Manager''. The main questions asked, presented in Table \ref{tab:survey_firewall_vendors}, revolved around the capabilities offered by their firewall products with regard to handling QUIC traffic. Specifically, the survey sought to determine whether the vendors \1 recommend blocking QUIC entirely, \2 include information about the QUIC protocol into their firewall state tables, \3 can differentiate QUIC traffic from other types of traffic, and \4 can perform basic filtering of QUIC traffic. Additionally, it was assessed whether the firewalls \5 currently offer functionality to decrypt QUIC traffic, which is similar to HTTPS decryption on a technical level, and \6 can perform HTTP/3 DPI. Full decryption of QUIC traffic is a pre-requisite for performing HTTP/3 DPI, which three out of five vendors are currently offering as part of their product suite. DPI refers to analyzing the contents of the packet after decrypting it, although it is also sometimes referred to as header analysis of the unencrypted QUIC handshake portions, such as the TLS Server Name Indication (SNI) field. Full decryption means that the firewall performs a man-in-the-middle inspection on QUIC traffic, making all embedded contents (\eg HTTP/3) readable to the firewall and enabling fine-grained content filtering. No vendor mentioned that their firewall solutions can reliably identify QUIC connection migrations -- neither client-side nor server-side \7.

Almost all surveyed vendors are \8 planning on developing or currently actively developing new QUIC-related features. However, as of now, they still recommend that their clients block the QUIC protocol entirely. Interestingly, all participants shared that QUIC traffic analysis features are rarely requested by clients, which suggests that there may be a lack of awareness or understanding about the importance and benefits of QUIC traffic analysis. Additionally, the firewall vendors were asked for their perspective on how their competitors are dealing with the increase in QUIC traffic. The general consensus among four out of five participants was that QUIC traffic analysis seems to be, as expected, an industry-wide challenge \9.

\section{Discussion}\label{sec:discussion}

As of today, modern firewalls are not tracking QUIC connections in state tables, meaning that state tables have to treat every outgoing UDP packet as a new connection attempt \cite{gbur2021}. Firewalls generally operate under the assumption that outgoing traffic is safe, as it originates from within the trusted network. When an outgoing UDP packet arrives at the firewall, the firewall checks whether the state table contains an existing entry corresponding to the packet's 5-tuple. If a matching entry is found, the packet is processed according to the pre-established rules for that connection. If no matching entry exists, the firewall may create a new entry in the state table or take other actions based on its configuration.

Although there are ways to perform stateful treatment of QUIC traffic based on its few observable features, such stateful treatment requires trade-offs with the confidentiality and censorship-resistance of the protocol. RFC 9312 \cite{rfc9312} discusses the manageability of the protocol and analyzes ways to perform stateful treatment of QUIC traffic. Apart from observing the cleartext parts of the handshake and implementing custom QUIC extensions that unconceal more information, there are limited options available to reliably track connections. Using the CID as a stateful identifier is not possible, since the CIDs can be renegotiated at any time within the encrypted channel. Stateful firewalls cannot even rely on the detection of end-of-flow signals to terminate a connection, as end-of-flow signals are not visible to an on-path observer \cite{rfc9312}. Therefore, a QUIC-aware firewall would have to rely on timer-based state removals.

In the context of the proposed data exfiltration method, this means that, as of today, most enterprise-level network-based firewalls cannot detect a potential QUIC connection migration, nor can they differentiate between migrations and request forgery attacks.

\subsection{Mitigation Strategies}

\noindent
The risk of the proposed attack can be partly mitigated through the following countermeasures: 

\begin{itemize}
    \item A client may disable QUIC transport parameters, like active connection migration using the \texttt{disable\_active\_migration (0x0c)} flag, or remove the server \texttt{preferred\_address} field from the QUIC implementation. Any connection migration attempt can then be flagged as anomalous activity.
    \item A modification to QUIC's implementation so that the \newline \texttt{preferred\_address} transport parameter field is part of the unencrypted QUIC handshake. As a result, custom middlebox software can be developed that can recognize connection migration attempts by monitoring QUIC packets for changes in their source or destination IP addresses and/or ports. These changes can then be checked against the contents of the \texttt{preferred\_address} parameter and domain registration records of the new IP addresses to verify their legitimacy.
    \item Another approach is full decryption of QUIC traffic on a firewall level. Although not recommended for privacy reasons, as it defeats the purpose of the QUIC protocol, it allows for in-depth HTTP/3 inspection. Even in such a case, firewall vendors would still need to implement the aforementioned custom detection mechanisms to verify that the \texttt{preferred\_address} address field contains a legitimate address.
\end{itemize}
\section{Related Work}\label{sec:related_work}

\subsection{QUIC Request Forgery Attacks}

QUIC traffic is inherently difficult to differentiate from ``normal'' UDP traffic. The encryption of QUIC headers and payloads obfuscates the traffic, which prevents easy inspection and classification. Even advanced network sniffing tools such as Wireshark cannot reliably detect QUIC if the handshake phase has not been observed. In addition to obfuscation challenges, there are various attacks targeting the QUIC protocol, for instance, request forgery attacks.

RFC 9000 \cite{rfc9000} describes different types of request forgery attacks, including ``Request Forgery with Client Initial Packets'', ``Request Forgery with Preferred Addresses'', ``Request Forgery with Spoofed Migration'', and ``Request Forgery with Version Negotiation''. Although \cite{rfc9000} mentions a request forgery attack on the DCID field in packets sent to a preferred address, it fails to point out that a packet's destination IP may be spoofed to redirect traffic to a malicious IP address. The RFC suggests no specific countermeasures beyond generic security recommendations.

Gbur and Tschorsch \cite{gbur2023} performed an analysis of the feasibility of client-side request forgery attacks. They focused on client-side Server Initial Request Forgery (SIRF), Version Negotiation Request Forgery (VNRF), and Connection Migration Request Forgery (CMRF). Their CMRF analysis only encompassed forging client-side connection migration events, which aim to trick a legitimate QUIC server into sending a QUIC packet to a spoofed address. It did not cover the forging of server-side connection migration events.

One example of spoofing connection migration events with benevolent intent is MIMIQ \cite{govil2020}, a privacy-enhancing system that aims to prevent traffic analysis by a middlebox. The system allows clients to maintain anonymity by frequently rotating their source IP address without disrupting connections. It prevents adversaries from identifying the client or associating multiple flows with it. Connections are broken into smaller flows and migration times are strategically chosen, making it harder for adversaries to analyze traffic and gather information about a particular client.

\subsection{Data Exfiltration}

The field of data exfiltration is vast, encompassing various attack vectors such as exfiltration over web services, physical media, network media, and alternative [network] protocols \cite{mitre}. We therefore limited the review to papers that developed attacks related to the MITRE ATT\&CK framework technique ``Exfiltration over Alternative Protocol'' \cite{mitre}, which includes all network protocols not being used as the main C2 channel.

There have been efforts in QUIC-based data exfiltration, such as the prototype presented in \cite{quicexfilproto}, which embeds data within a legitimate QUIC connection. However, this approach does not attempt to hide the exfiltration, leaves a noticeable fingerprint due to the handshake, and does not adjust packet features like payload length.

Sudhan and Kulkarni \cite{kulkarni2024} introduced a method to establish a covert channel between two QUIC endpoints using the latency spin bit. The spin bit is an optional QUIC protocol feature that allows for passive on-path network latency monitoring. The proposed method requires two QUIC endpoints to establish a legitimate connection, making it less feasible for malicious data exfiltration attempts. Furthermore, only one bit of data can be exchanged per QUIC packet, which makes the method impractical for exfiltrating large amounts of data.

Zhan et al. \cite{zhan2022} proposed a method to detect DNS-over-HTTPS (DoH) based data exfiltration by analyzing TLS-fingerprints and training Boosted Decision Trees, Random Forest, and Logistic Regression classifiers on flow-based features, achieving detection accuracies of over 99\%.

Vaccari et al. \cite{vaccari2021} exploit the Message Queue Telemetry Transport (MQTT) protocol, commonly used within IoT networks, to exfiltrate sensitive data from a private network. Their method successfully exfiltrated payloads up to 3000 bytes over the MQTT protocol, and simultaneously, they were able to achieve detection accuracies of up to 99\% using Random Forest classifiers.

Klein \cite{klein2022} introduced a data exfiltration method that exploits stateful IPv4 IDs, TCP ISNs and IPv6 flow labels on popular server operating systems. The study demonstrated the feasibility of exfiltrating data from \textit{firewalled} networks using the global protocol states to establish covert channels. In addition, it explored cross-protocol attacks and measured exfiltration bandwidth, which was sufficiently high to extract secret key material within a few hours.

\section{Conclusion}\label{sec:conclusion}

This paper analyzes the feasibility of covert data exfiltration attacks using the QUIC transport protocol. We find that adversaries can make use of QUIC's server-side connection migration feature to exfiltrate data from a trusted network to a target server. We show that, because of the inherent traits of the QUIC protocol, QUIC-based data exfiltration techniques are difficult to differentiate from normal protocol behaviour, and not even custom anomaly detection classifiers are able to detect such data exfiltration attempts. Some mitigation strategies include outright disabling server-side connection migration, sending the \texttt{preferred\_address} parameter as part of the unencrypted handshake, or implementing custom firewall software that checks the \texttt{preferred\_address} parameter against domain registration entries. From the lack of QUIC traffic analysis capabilities offered by leading firewall vendors, one can infer that, as of today, firewalls cannot effectively handle the complexities of QUIC. Our contribution lies not in demonstrating the success of ML-based detection, but rather in revealing the limitations of current ML methods against advanced mimicking attacks. Future work may include developing heuristics-based QUIC traffic inspectors that can be deployed on middleboxes.

\begin{acks}

This work was partially supported by (a) the University of Zürich UZH, Switzerland, and (b) the Horizon Europe Framework Program's project Certify, Grant Agreement No. 101069471, funded by the Swiss State Secretariat for Education, Research, and Innovation SERI, under Contract No. 22.00165.

\end{acks}

\newpage

\balance
\bibliographystyle{ACM-Reference-Format}
\bibliography{base}


\begin{thebibliography}{49}


\ifx \showCODEN    \undefined \def \showCODEN     #1{\unskip}     \fi
\ifx \showDOI      \undefined \def \showDOI       #1{#1}\fi
\ifx \showISBNx    \undefined \def \showISBNx     #1{\unskip}     \fi
\ifx \showISBNxiii \undefined \def \showISBNxiii  #1{\unskip}     \fi
\ifx \showISSN     \undefined \def \showISSN      #1{\unskip}     \fi
\ifx \showLCCN     \undefined \def \showLCCN      #1{\unskip}     \fi
\ifx \shownote     \undefined \def \shownote      #1{#1}          \fi
\ifx \showarticletitle \undefined \def \showarticletitle #1{#1}   \fi
\ifx \showURL      \undefined \def \showURL       {\relax}        \fi
\providecommand\bibfield[2]{#2}
\providecommand\bibinfo[2]{#2}
\providecommand\natexlab[1]{#1}
\providecommand\showeprint[2][]{arXiv:#2}

\bibitem[accetto(2024)]%
        {dockercontainer}
\bibfield{author}{\bibinfo{person}{accetto}.} \bibinfo{year}{2024}\natexlab{}.
\newblock \bibinfo{booktitle}{\emph{Headless Ubuntu/Xfce container with
  VNC/noVNC and Firefox (G2).}}
\newblock
\urldef\tempurl%
\url{https://hub.docker.com/r/accetto/xubuntu-vnc-novnc-firefox}
\showURL{%
\tempurl}


\bibitem[{Alibaba}(2024)]%
        {xquic}
\bibfield{author}{\bibinfo{person}{{Alibaba}}.}
  \bibinfo{year}{2024}\natexlab{}.
\newblock \bibinfo{booktitle}{\emph{xquic}}.
\newblock
\urldef\tempurl%
\url{https://github.com/alibaba/xquic}
\showURL{%
\tempurl}


\bibitem[Belson and Pardue(2023)]%
        {cloudflarehttp3}
\bibfield{author}{\bibinfo{person}{David Belson} {and} \bibinfo{person}{Lucas
  Pardue}.} \bibinfo{year}{2023}\natexlab{}.
\newblock \bibinfo{booktitle}{\emph{Examining HTTP/3 usage one year on}}.
\newblock
\urldef\tempurl%
\url{https://blog.cloudflare.com/http3-usage-one-year-on}
\showURL{%
\tempurl}


\bibitem[Bishop(2022)]%
        {rfc9114}
\bibfield{author}{\bibinfo{person}{M. Bishop}.}
  \bibinfo{year}{2022}\natexlab{}.
\newblock \bibinfo{booktitle}{\emph{HTTP/3}}.
\newblock \bibinfo{type}{RFC} 9114. \bibinfo{institution}{IETF}.
\newblock
\showISSN{2070-1721}
\urldef\tempurl%
\url{https://www.ietf.org/rfc/rfc9114.txt}
\showURL{%
\tempurl}


\bibitem[Buchet and Pelsser(2024)]%
        {buchet2024}
\bibfield{author}{\bibinfo{person}{Aur{\'e}lien Buchet} {and}
  \bibinfo{person}{Cristel Pelsser}.} \bibinfo{year}{2024}\natexlab{}.
\newblock \showarticletitle{An Analysis of QUIC Connection Migration in the
  Wild}.
\newblock \bibinfo{journal}{\emph{arXiv preprint arXiv:2410.06066}}
  (\bibinfo{year}{2024}).
\newblock


\bibitem[Cisco(2023)]%
        {ciscoevi}
\bibfield{author}{\bibinfo{person}{Cisco}.} \bibinfo{year}{2023}\natexlab{}.
\newblock \bibinfo{booktitle}{\emph{Encrypted Visibility Engine: An overview of
  Cisco Secure Firewall's Encrypted Visibility Engine (EVE)}}.
\newblock
\urldef\tempurl%
\url{https://secure.cisco.com/secure-firewall/v7.3/docs/encrypted-visibility-engine-73}
\showURL{%
\tempurl}


\bibitem[{Cloudflare}(2024)]%
        {cloudflare-quiche}
\bibfield{author}{\bibinfo{person}{{Cloudflare}}.}
  \bibinfo{year}{2024}\natexlab{}.
\newblock \bibinfo{booktitle}{\emph{quiche}}.
\newblock
\urldef\tempurl%
\url{https://github.com/cloudflare/quiche/blob/master/quiche/src/lib.rs}
\showURL{%
\tempurl}


\bibitem[{devsisters}(2024)]%
        {libquic}
\bibfield{author}{\bibinfo{person}{{devsisters}}.}
  \bibinfo{year}{2024}\natexlab{}.
\newblock \bibinfo{booktitle}{\emph{libquic}}.
\newblock
\urldef\tempurl%
\url{https://github.com/devsisters/libquic}
\showURL{%
\tempurl}


\bibitem[{ebfull and Kozlowski, Wojciech}(2024)]%
        {pcap}
\bibfield{author}{\bibinfo{person}{{ebfull and Kozlowski, Wojciech}}.}
  \bibinfo{year}{2024}\natexlab{}.
\newblock \bibinfo{booktitle}{\emph{pcap: A packet capture API around
  pcap/wpcap}}.
\newblock
\urldef\tempurl%
\url{https://crates.io/crates/pcap}
\showURL{%
\tempurl}


\bibitem[Gartner(2024)]%
        {gartner}
\bibfield{author}{\bibinfo{person}{Gartner}.} \bibinfo{year}{2024}\natexlab{}.
\newblock \bibinfo{booktitle}{\emph{Network Firewalls Reviews and Ratings}}.
\newblock
\urldef\tempurl%
\url{https://www.gartner.com/reviews/market/network-firewalls}
\showURL{%
\tempurl}


\bibitem[Gbur and Tschorsch(2021)]%
        {gbur2021}
\bibfield{author}{\bibinfo{person}{Konrad~Yuri Gbur} {and}
  \bibinfo{person}{Florian Tschorsch}.} \bibinfo{year}{2021}\natexlab{}.
\newblock \showarticletitle{A quic (k) way through your firewall?}
\newblock \bibinfo{journal}{\emph{arXiv preprint arXiv:2107.05939}}
  (\bibinfo{year}{2021}).
\newblock


\bibitem[Gbur and Tschorsch(2023)]%
        {gbur2023}
\bibfield{author}{\bibinfo{person}{Konrad~Yuri Gbur} {and}
  \bibinfo{person}{Florian Tschorsch}.} \bibinfo{year}{2023}\natexlab{}.
\newblock \showarticletitle{QUICforge: Client-side Request Forgery in QUIC}. In
  \bibinfo{booktitle}{\emph{Network and Distributed System Security (NDSS)
  Symposium}}.
\newblock


\bibitem[Goldberg et~al\mbox{.}(2020)]%
        {goldberg2020}
\bibfield{author}{\bibinfo{person}{I. Goldberg}, \bibinfo{person}{T. Wang},
  {and} \bibinfo{person}{C.A. Wood}.} \bibinfo{year}{2020}\natexlab{}.
\newblock \bibinfo{booktitle}{\emph{Network-Based Website Fingerprinting}}.
\newblock \bibinfo{type}{{T}echnical {R}eport}. \bibinfo{institution}{pearg}.
\newblock
\urldef\tempurl%
\url{https://www.ietf.org/archive/id/draft-irtf-pearg-website-fingerprinting-00.html}
\showURL{%
\tempurl}


\bibitem[{Google}(2024)]%
        {gquiche}
\bibfield{author}{\bibinfo{person}{{Google}}.} \bibinfo{year}{2024}\natexlab{}.
\newblock \bibinfo{booktitle}{\emph{Google quiche}}.
\newblock
\urldef\tempurl%
\url{https://github.com/google/quiche}
\showURL{%
\tempurl}


\bibitem[Govil et~al\mbox{.}(2020)]%
        {govil2020}
\bibfield{author}{\bibinfo{person}{Yashodhar Govil}, \bibinfo{person}{Liang
  Wang}, {and} \bibinfo{person}{Jennifer Rexford}.}
  \bibinfo{year}{2020}\natexlab{}.
\newblock \showarticletitle{$\{$MIMIQ$\}$: Masking $\{$IPs$\}$ with Migration
  in $\{$QUIC$\}$}. In \bibinfo{booktitle}{\emph{10th USENIX Workshop on Free
  and Open Communications on the Internet (FOCI 20)}}.
\newblock


\bibitem[{HAProxy}(2024)]%
        {haproxy}
\bibfield{author}{\bibinfo{person}{{HAProxy}}.}
  \bibinfo{year}{2024}\natexlab{}.
\newblock \bibinfo{booktitle}{\emph{haproxy}}.
\newblock
\urldef\tempurl%
\url{https://www.haproxy.org/}
\showURL{%
\tempurl}


\bibitem[Huitema et~al\mbox{.}(2024)]%
        {rfc9250}
\bibfield{author}{\bibinfo{person}{Christian Huitema}, \bibinfo{person}{Sara
  Dickinson}, {and} \bibinfo{person}{Allison Mankin}.}
  \bibinfo{year}{2024}\natexlab{}.
\newblock \bibinfo{booktitle}{}.
\newblock \bibinfo{type}{RFC} 9250. \bibinfo{institution}{IETF}.
\newblock
\showISSN{2070-1721}
\urldef\tempurl%
\url{https://www.rfc-editor.org/rfc/rfc9250.txt}
\showURL{%
\tempurl}


\bibitem[IBM(2024)]%
        {ibm}
\bibfield{author}{\bibinfo{person}{IBM}.} \bibinfo{year}{2024}\natexlab{}.
\newblock \bibinfo{booktitle}{\emph{Cost of a Data Breach Report 2024}}.
\newblock
\urldef\tempurl%
\url{https://www.ibm.com/reports/data-breach}
\showURL{%
\tempurl}


\bibitem[Iyengar and Thomson(2021)]%
        {rfc9000}
\bibfield{author}{\bibinfo{person}{Jana Iyengar} {and} \bibinfo{person}{Martin
  Thomson}.} \bibinfo{year}{2021}\natexlab{}.
\newblock \bibinfo{booktitle}{\emph{QUIC: A UDP-Based Multiplexed and Secure
  Transport}}.
\newblock \bibinfo{type}{RFC} 9000. \bibinfo{institution}{IETF}.
\newblock
\showISSN{2070-1721}
\urldef\tempurl%
\url{https://www.ietf.org/rfc/rfc9000.txt}
\showURL{%
\tempurl}


\bibitem[{Karimi, Adel}(2019)]%
        {fatt}
\bibfield{author}{\bibinfo{person}{{Karimi, Adel}}.}
  \bibinfo{year}{2019}\natexlab{}.
\newblock \bibinfo{booktitle}{\emph{FATT: fingerprint all the things!}}
\newblock
\urldef\tempurl%
\url{https://github.com/0x4D31/fatt}
\showURL{%
\tempurl}


\bibitem[Klein(2022)]%
        {klein2022}
\bibfield{author}{\bibinfo{person}{Amit Klein}.}
  \bibinfo{year}{2022}\natexlab{}.
\newblock \showarticletitle{Subverting Stateful Firewalls with Protocol
  States}. In \bibinfo{booktitle}{\emph{Network and Distributed System Security
  (NDSS) Symposium}}.
\newblock


\bibitem[Kühlewind and Trammell(2022)]%
        {rfc9312}
\bibfield{author}{\bibinfo{person}{M. Kühlewind} {and} \bibinfo{person}{B.
  Trammell}.} \bibinfo{year}{2022}\natexlab{}.
\newblock \bibinfo{booktitle}{\emph{Manageability of the QUIC Transport
  Protocol}}.
\newblock \bibinfo{type}{RFC} 9312. \bibinfo{institution}{IETF}.
\newblock
\showISSN{2070-1721}
\urldef\tempurl%
\url{https://www.ietf.org/rfc/rfc9312.txt}
\showURL{%
\tempurl}


\bibitem[{Lainé, Jeremy }(2024)]%
        {aioquic}
\bibfield{author}{\bibinfo{person}{{Lainé, Jeremy }}.}
  \bibinfo{year}{2024}\natexlab{}.
\newblock \bibinfo{booktitle}{\emph{aioquic}}.
\newblock
\urldef\tempurl%
\url{https://github.com/aiortc/aioquic}
\showURL{%
\tempurl}


\bibitem[Langley et~al\mbox{.}(2017)]%
        {langley2017}
\bibfield{author}{\bibinfo{person}{Adam Langley}, \bibinfo{person}{Alistair
  Riddoch}, \bibinfo{person}{Alyssa Wilk}, \bibinfo{person}{Antonio Vicente},
  \bibinfo{person}{Charles Krasic}, \bibinfo{person}{Dan Zhang},
  \bibinfo{person}{Fan Yang}, \bibinfo{person}{Fedor Kouranov},
  \bibinfo{person}{Ian Swett}, \bibinfo{person}{Janardhan Iyengar},
  {et~al\mbox{.}}} \bibinfo{year}{2017}\natexlab{}.
\newblock \showarticletitle{The quic transport protocol: Design and
  internet-scale deployment}. In \bibinfo{booktitle}{\emph{Proceedings of the
  conference of the ACM special interest group on data communication}}.
  \bibinfo{pages}{183--196}.
\newblock


\bibitem[Liu et~al\mbox{.}(2024)]%
        {liu2024}
\bibfield{author}{\bibinfo{person}{Yanmei Liu}, \bibinfo{person}{Yunfei Ma},
  \bibinfo{person}{Quentin~De Coninck}, \bibinfo{person}{Olivier Bonaventure},
  \bibinfo{person}{Christian Huitema}, {and} \bibinfo{person}{Mirja
  Kühlewind}.} \bibinfo{year}{2024}\natexlab{}.
\newblock \bibinfo{booktitle}{\emph{Multipath Extension for QUIC}}.
\newblock \bibinfo{type}{Work in Progress}. \bibinfo{institution}{IETF}.
\newblock
\urldef\tempurl%
\url{https://datatracker.ietf.org/doc/draft-ietf-quic-multipath/}
\showURL{%
\tempurl}


\bibitem[Lizhuang et~al\mbox{.}(2020)]%
        {tan2020}
\bibfield{author}{\bibinfo{person}{Tan Lizhuang}, \bibinfo{person}{Gao
  Xiaochuan}, \bibinfo{person}{Su Wei}, \bibinfo{person}{Li Na}, {and}
  \bibinfo{person}{Zhang Wei}.} \bibinfo{year}{2020}\natexlab{}.
\newblock \bibinfo{booktitle}{\emph{Connection Migration in QUIC}}.
\newblock \bibinfo{type}{Work in Progress}. \bibinfo{institution}{IETF}.
\newblock
\urldef\tempurl%
\url{https://datatracker.ietf.org/doc/html/draft-tan-quic-connection-migration-00}
\showURL{%
\tempurl}


\bibitem[Lu et~al\mbox{.}(2021)]%
        {lu2021}
\bibfield{author}{\bibinfo{person}{Chaoyi Lu}, \bibinfo{person}{Baojun Liu},
  \bibinfo{person}{Yiming Zhang}, \bibinfo{person}{Zhou Li},
  \bibinfo{person}{Fenglu Zhang}, \bibinfo{person}{Haixin Duan},
  \bibinfo{person}{Ying Liu}, \bibinfo{person}{Joann~Qiongna Chen},
  \bibinfo{person}{Jinjin Liang}, \bibinfo{person}{Zaifeng Zhang},
  {et~al\mbox{.}}} \bibinfo{year}{2021}\natexlab{}.
\newblock \showarticletitle{From WHOIS to WHOWAS: A Large-Scale Measurement
  Study of Domain Registration Privacy under the GDPR}. In
  \bibinfo{booktitle}{\emph{Network and Distributed System Security (NDSS)
  Symposium}}.
\newblock


\bibitem[Lundberg and Lee(2017)]%
        {NIPS2017_7062}
\bibfield{author}{\bibinfo{person}{Scott~M Lundberg} {and}
  \bibinfo{person}{Su-In Lee}.} \bibinfo{year}{2017}\natexlab{}.
\newblock \showarticletitle{A Unified Approach to Interpreting Model
  Predictions}.
\newblock In \bibinfo{booktitle}{\emph{Advances in Neural Information
  Processing Systems 30}}, \bibfield{editor}{\bibinfo{person}{I.~Guyon},
  \bibinfo{person}{U.~V. Luxburg}, \bibinfo{person}{S.~Bengio},
  \bibinfo{person}{H.~Wallach}, \bibinfo{person}{R.~Fergus},
  \bibinfo{person}{S.~Vishwanathan}, {and} \bibinfo{person}{R.~Garnett}}
  (Eds.). \bibinfo{publisher}{Curran Associates, Inc.},
  \bibinfo{pages}{4765--4774}.
\newblock
\urldef\tempurl%
\url{http://papers.nips.cc/paper/7062-a-unified-approach-to-interpreting-model-predictions.pdf}
\showURL{%
\tempurl}


\bibitem[Martini and ten Oever(2019)]%
        {martini2019quic}
\bibfield{author}{\bibinfo{person}{Beatrice Martini} {and}
  \bibinfo{person}{Niels ten Oever}.} \bibinfo{year}{2019}\natexlab{}.
\newblock \bibinfo{booktitle}{\emph{QUIC Human Rights Review}}.
\newblock \bibinfo{type}{{T}echnical {R}eport}. \bibinfo{institution}{HRPC
  Human Rights Review Team}.
\newblock
\urldef\tempurl%
\url{https://datatracker.ietf.org/meeting/102/materials/slides-102-hrpc-slides-hrreveiw-quic-00}
\showURL{%
\tempurl}


\bibitem[{McGrew, David and Enright, Brandon and Anderson, Blake and Messenger,
  Lucas and Weller, Adam and Chi, Andrew and Acharya, Shekhar and Antonyk,
  Anastasiia-Mariia and Stepanov, Oleksandr and Viswanathan, Vigneshwari and
  Raj, Apoorv}(2020)]%
        {mercury}
\bibfield{author}{\bibinfo{person}{{McGrew, David and Enright, Brandon and
  Anderson, Blake and Messenger, Lucas and Weller, Adam and Chi, Andrew and
  Acharya, Shekhar and Antonyk, Anastasiia-Mariia and Stepanov, Oleksandr and
  Viswanathan, Vigneshwari and Raj, Apoorv}}.} \bibinfo{year}{2020}\natexlab{}.
\newblock \bibinfo{booktitle}{\emph{Cisco Mercury: network metadata capture and
  analysis}}.
\newblock
\urldef\tempurl%
\url{https://github.com/cisco/mercury}
\showURL{%
\tempurl}


\bibitem[MITRE(2024)]%
        {mitre}
\bibfield{author}{\bibinfo{person}{MITRE}.} \bibinfo{year}{2024}\natexlab{}.
\newblock \bibinfo{title}{MITRE ATT\&CK Framework. Exfiltration Over
  Alternative Protocol.}
\newblock
\newblock
\urldef\tempurl%
\url{https://attack.mitre.org/techniques/T1048/}
\showURL{%
\tempurl}


\bibitem[Piraux and Bonaventure(2024)]%
        {piraux2024}
\bibfield{author}{\bibinfo{person}{Maxime Piraux} {and}
  \bibinfo{person}{Olivier Bonaventure}.} \bibinfo{year}{2024}\natexlab{}.
\newblock \bibinfo{booktitle}{\emph{Additional addresses for QUIC}}.
\newblock \bibinfo{type}{Work in Progress}. \bibinfo{institution}{IETF}.
\newblock
\urldef\tempurl%
\url{https://datatracker.ietf.org/doc/draft-piraux-quic-additional-addresses/}
\showURL{%
\tempurl}


\bibitem[Pornin(2021)]%
        {rfc8999}
\bibfield{author}{\bibinfo{person}{T. Pornin}.}
  \bibinfo{year}{2021}\natexlab{}.
\newblock \bibinfo{booktitle}{\emph{Version-Independent Properties of QUIC}}.
\newblock \bibinfo{type}{RFC} 8999. \bibinfo{institution}{IETF}.
\newblock
\showISSN{2070-1721}
\urldef\tempurl%
\url{https://www.ietf.org/rfc/rfc8999.txt}
\showURL{%
\tempurl}


\bibitem[{private-octopus}(2024)]%
        {picoquic}
\bibfield{author}{\bibinfo{person}{{private-octopus}}.}
  \bibinfo{year}{2024}\natexlab{}.
\newblock \bibinfo{booktitle}{\emph{picoquic}}.
\newblock
\urldef\tempurl%
\url{https://github.com/private-octopus/picoquic}
\showURL{%
\tempurl}


\bibitem[Puliafito et~al\mbox{.}(2022)]%
        {puliafito2022}
\bibfield{author}{\bibinfo{person}{Carlo Puliafito}, \bibinfo{person}{Luca
  Conforti}, \bibinfo{person}{Antonio Virdis}, {and} \bibinfo{person}{Enzo
  Mingozzi}.} \bibinfo{year}{2022}\natexlab{}.
\newblock \showarticletitle{Server-side QUIC connection migration to support
  microservice deployment at the edge}.
\newblock \bibinfo{journal}{\emph{Pervasive and mobile computing}}
  \bibinfo{volume}{83} (\bibinfo{year}{2022}), \bibinfo{pages}{101580}.
\newblock


\bibitem[{Puliafito, Carlo and Conforti, Luca and Virdis, Antonio and Mingozzi,
  Enzo}(2022)]%
        {aioquic_pisa}
\bibfield{author}{\bibinfo{person}{{Puliafito, Carlo and Conforti, Luca and
  Virdis, Antonio and Mingozzi, Enzo}}.} \bibinfo{year}{2022}\natexlab{}.
\newblock \bibinfo{booktitle}{\emph{Server-side QUIC connection migration to
  support microservice deployment at the edge}}.
\newblock
\urldef\tempurl%
\url{https://github.com/kruviser/aioquic-explicit\_UniPisa}
\showURL{%
\tempurl}


\bibitem[{quic-go}(2024)]%
        {quicgo}
\bibfield{author}{\bibinfo{person}{{quic-go}}.}
  \bibinfo{year}{2024}\natexlab{}.
\newblock \bibinfo{booktitle}{\emph{quic-ggo}}.
\newblock
\urldef\tempurl%
\url{https://github.com/quic-go/quic-go}
\showURL{%
\tempurl}


\bibitem[{quinn-rs}(2024)]%
        {quinn}
\bibfield{author}{\bibinfo{person}{{quinn-rs}}.}
  \bibinfo{year}{2024}\natexlab{}.
\newblock \bibinfo{booktitle}{\emph{quinn}}.
\newblock
\urldef\tempurl%
\url{https://github.com/quinn-rs/quinn}
\showURL{%
\tempurl}


\bibitem[Roskind(2012)]%
        {roskind2012}
\bibfield{author}{\bibinfo{person}{Jim Roskind}.}
  \bibinfo{year}{2012}\natexlab{}.
\newblock \bibinfo{booktitle}{\emph{Quick UDP internet connections: Multiplexed
  stream transport over UDP}}.
\newblock
\urldef\tempurl%
\url{https://www.ietf.org/proceedings/88/slides/slides-88-tsvarea-10.pdf}
\showURL{%
\tempurl}


\bibitem[{Schmid, Julian}(2024)]%
        {etherparse}
\bibfield{author}{\bibinfo{person}{{Schmid, Julian}}.}
  \bibinfo{year}{2024}\natexlab{}.
\newblock \bibinfo{booktitle}{\emph{etherparse}}.
\newblock
\urldef\tempurl%
\url{https://crates.io/crates/etherparse}
\showURL{%
\tempurl}


\bibitem[Shi(2019)]%
        {shi2019}
\bibfield{author}{\bibinfo{person}{Cherie Shi}.}
  \bibinfo{year}{2019}\natexlab{}.
\newblock \bibinfo{booktitle}{\emph{QUIC Connection Migration}}.
\newblock \bibinfo{type}{{T}echnical {R}eport}. \bibinfo{institution}{IETF}.
\newblock
\showISSN{104}
\urldef\tempurl%
\url{https://datatracker.ietf.org/doc/slides-104-maprg-quic-connection-migration-cherie-shi/}
\showURL{%
\tempurl}


\bibitem[Stewart et~al\mbox{.}(2022)]%
        {rfc9260}
\bibfield{author}{\bibinfo{person}{Randall~R. Stewart},
  \bibinfo{person}{Michael Tüxen}, {and} \bibinfo{person}{karen Nielsen}.}
  \bibinfo{year}{2022}\natexlab{}.
\newblock \bibinfo{title}{{Stream Control Transmission Protocol}}.
\newblock \bibinfo{howpublished}{RFC 9260}.
\newblock
\urldef\tempurl%
\url{https://doi.org/10.17487/RFC9260}
\showDOI{\tempurl}


\bibitem[Sudhan~S. and Kulkarni(2024)]%
        {kulkarni2024}
\bibfield{author}{\bibinfo{person}{H.~H. Sudhan~S.} {and}
  \bibinfo{person}{Sameer~G. Kulkarni}.} \bibinfo{year}{2024}\natexlab{}.
\newblock \showarticletitle{Security and Service Vulnerabilities with HTTP/3}.
  In \bibinfo{booktitle}{\emph{2024 16th International Conference on
  COMmunication Systems \& NETworkS (COMSNETS)}}. IEEE,
  \bibinfo{pages}{55--60}.
\newblock


\bibitem[Vaccari et~al\mbox{.}(2021)]%
        {vaccari2021}
\bibfield{author}{\bibinfo{person}{Ivan Vaccari}, \bibinfo{person}{Sara
  Narteni}, \bibinfo{person}{Maurizio Aiello}, \bibinfo{person}{Maurizio
  Mongelli}, {and} \bibinfo{person}{Enrico Cambiaso}.}
  \bibinfo{year}{2021}\natexlab{}.
\newblock \showarticletitle{Exploiting Internet of Things protocols for
  malicious data exfiltration activities}.
\newblock \bibinfo{journal}{\emph{IEEE Access}}  \bibinfo{volume}{9}
  (\bibinfo{year}{2021}), \bibinfo{pages}{104261--104280}.
\newblock


\bibitem[Wang et~al\mbox{.}(2022)]%
        {wang2022}
\bibfield{author}{\bibinfo{person}{Mona Wang}, \bibinfo{person}{Anunay
  Kulshrestha}, \bibinfo{person}{Liang Wang}, {and} \bibinfo{person}{Prateek
  Mittal}.} \bibinfo{year}{2022}\natexlab{}.
\newblock \showarticletitle{Leveraging strategic connection migration-powered
  traffic splitting for privacy}.
\newblock \bibinfo{journal}{\emph{arXiv preprint arXiv:2205.03326}}
  (\bibinfo{year}{2022}).
\newblock


\bibitem[{Wireshark Foundation}(2024)]%
        {wireshark}
\bibfield{author}{\bibinfo{person}{{Wireshark Foundation}}.}
  \bibinfo{year}{2024}\natexlab{}.
\newblock \bibinfo{booktitle}{\emph{Wireshark}}.
\newblock
\urldef\tempurl%
\url{https://www.wireshark.org}
\showURL{%
\tempurl}


\bibitem[{Yamamoto, Kazu}(2024)]%
        {haskell_quic}
\bibfield{author}{\bibinfo{person}{{Yamamoto, Kazu}}.}
  \bibinfo{year}{2024}\natexlab{}.
\newblock \bibinfo{booktitle}{\emph{IETF QUIC implementation in Haskell}}.
\newblock
\urldef\tempurl%
\url{https://hackage.haskell.org/package/quic}
\showURL{%
\tempurl}


\bibitem[ytisf(2024)]%
        {quicexfilproto}
\bibfield{author}{\bibinfo{person}{ytisf}.} \bibinfo{year}{2024}\natexlab{}.
\newblock \bibinfo{booktitle}{\emph{PyExfil: Stress Testing Detection \&
  Creativity}}.
\newblock
\urldef\tempurl%
\url{https://github.com/ytisf/PyExfil/blob/master/USAGE.md}
\showURL{%
\tempurl}


\bibitem[Zhan et~al\mbox{.}(2022)]%
        {zhan2022}
\bibfield{author}{\bibinfo{person}{Mengqi Zhan}, \bibinfo{person}{Yang Li},
  \bibinfo{person}{Guangxi Yu}, \bibinfo{person}{Bo Li}, {and}
  \bibinfo{person}{Weiping Wang}.} \bibinfo{year}{2022}\natexlab{}.
\newblock \showarticletitle{Detecting DNS over HTTPS based data exfiltration}.
\newblock \bibinfo{journal}{\emph{Computer Networks}}  \bibinfo{volume}{209}
  (\bibinfo{year}{2022}), \bibinfo{pages}{108919}.
\newblock


\end{thebibliography}

\noindent \small{\\All links above were last accessed on \today.}

\appendix

\section{Evaluation of Open-Source Implementations}

Since the proposed attack relies on client-initiated server-side migrations, it is vital to understand the actual adoption of this feature. Thus, this section reviews several open-source client- and server-side implementations of the QUIC protocol. Each solution (\cf \tablename{}~\ref{tab:static-analysis}) is statically analyzed to infer whether it supports server-side connection migrations and/or Multipath QUIC \cite{liu2024}. Furthermore, since libraries differ in terms of maturity and adoption, a non-exhaustive set of dependents of each library (\ie other libraries, clients, or applications) is enumerated.

Currently, 8 out of the 11 reviewed libraries implement the feature, whereas the remaining 3 only implement client-side connection migrations. For example, \textit{quic-go} is a widely-used library that does not actively support server-side connection migration. \textit{aioquic} parses the \texttt{preferred\_address} field, but does not support active migration. Similarly, \textit{Cloudflare quiche} does not support server-side connection migration. However, from the source code, it can be inferred that it is planned to implement the feature. On the other hand, several libraries already allow active server-side connection migrations and/or QUIC Multipath connections. For example, \textit{picoquic}, \textit{libquic}, \textit{quinn}, \textit{aioquic\_pisa}, \textit{Google quiche}, \textit{haproxy}, \textit{Haskell quic}, and Alibaba's \textit{xquic} support the feature. These implementations are used by several operating systems and user-space applications, including the Google Chrome browser.

Thus, based on these observations, it can be argued that server-side connection migration is not a hypothetical feature of QUIC but instead a relevant feature of the protocol. Therefore, the previously mentioned attack is exploiting a well-established feature for several implementations. However, it must be mentioned that there is no empirical evidence of the feature's prevalence in network traffic.

\begin{table}[t]
    \centering
    \caption{Overview of Statically Analyzed Implementations}
    \label{tab:static-analysis}
    \scalebox{0.9}{
    \begin{tabular}{@{}l|c p{3.75cm} @{}}
        \toprule
        \textbf{\textit{Implementation}} & \textbf{\textit{Implemented}} &  \textbf{\textit{Notable Dependents}} \\
        \midrule
        \textit{quic-go}~\cite{quicgo}  & \xmark   & cloudflared, caddy, syncthing   \\
        \textit{libquic}~\cite{libquic} & \checkmark  & goquic, chromium \\
        \textit{Cloudflare quiche}~\cite{cloudflare-quiche} & \xmark *  & Cloudflare, NGINX \\
        \textit{picoquic}~\cite{picoquic}  & \checkmark   &  Picotls, RIOT OS \\
        \textit{quinn}~\cite{quinn}  & \checkmark   &  h3-quinn, nestri, EasyTier \\
        \textit{aioquic}~\cite{aioquic}  & \xmark  & airbyte, mitmproxy, envoy\\
        \textit{aioquic\_pisa}~\cite{aioquic_pisa}  & \checkmark  & -- \\
        \textit{Google quiche}~\cite{gquiche}  & \checkmark   &  Google, Chromium\\
        \textit{haproxy}~\cite{haproxy}  &  \checkmark & Instagram, Airbnb, pfSense\\
        \textit{Haskell quic}~\cite{haskell_quic}  &  \checkmark & hprox, warp-quic, http3  \\
        \textit{Alibaba xquic}~\cite{xquic} &  \checkmark &  Taobao Mobile  \\
        \bottomrule
    \end{tabular}
    }
    \\ $*$Implementation of Server-Side Migration Planned \newline
\end{table}

\section{Wireshark Filters}

The following Wireshark filters were used to extract the datasets for training the anomaly detectors: \\

1) Matching all outgoing QUIC Protected Payload packets in the testbed: 
\lstset{language==[]}
\begin{lstlisting}[caption=Wireshark Filter Example 1, label=lst:wireshark_filter]
ip.src == 172.19.0.0/16 && quic && quic.header_form == "short header" && quic.header_form != "long header" && quic.dcid != ""
\end{lstlisting}

2) Matching all outgoing QUIC Protected Payload packets as well as all benign connection migration attempts (triggered using Cloudflare \textit{quiche}):

\lstset{language==[]}
\begin{lstlisting}[caption=Wireshark Filter Example 2, label=lst:wireshark_filter]
ip.src == 172.19.0.X && ((quic && quic.header_form == "short header" && quic.header_form != "long header" && quic.dcid != "") || (udp && udp.length == 1358))
\end{lstlisting}

\section{Additional Considerations}

\subsection{Potential Drawbacks of Establishing New QUIC Connections}

One of the primary challenges in executing a covert data exfiltration attack using the QUIC protocol arises from the transparency during the connection establishment phase. Middleboxes and fingerprinting tools typically filter connections based on the initial handshake packets. This initial packet exchange, which includes the ``Initial'' QUIC packet or potentially the ``0-RTT'' packet, serves as a clear indicator of new connection establishment. These packets contain cleartext details, such as the TLS Client Hello and TLS Server Hello messages, that can expose a certain fingerprint. Given the amount of metadata within the handshake packets, any attempts to establish a new connection to exfiltrate data would likely increase the visibility of the attack. In particular, fingerprinting tools (\cf Section \ref{subsec:fingerprinting}) filter exclusively based on handshakes and therefore increase the risk of the attack being identified and blocked at an early stage.

\subsection{Comparison with TLS and DNS-based Data Exfiltration}

Defense systems specifically look for TLS Client Hello or TCP SYN packets to identify connection establishments. TCP-based connections typically require a new handshake to establish a valid connection when an underlying IP address changes. Similarly to QUIC, which allows connection migrations that change the underlying IP address \textit{without} requiring a new handshake, there are further exceptions, such as the Stream Control Transmission Protocol (SCTP) \cite{rfc9260}, which can reconfigure IP addresses mid-connection using the \textit{Set Primary} instruction. However, SCTP packets in non-telecom networks are not as prevalent as QUIC traffic, and, as a result, QUIC-based data exfiltration that mimics legitimate connection migrations potentially poses a greater security concern.

DNS-based data exfiltration may raise suspicion when multiple standalone DNS queries are produced, and not followed by a TCP and/or TLS handshake after an IP address has been resolved. This makes high-throughput DNS-based data exfiltration practically impossible without attracting attention. QUIC, since it inherently anticipates changes in the underlying IP header, may make data exfiltration appear less anomalous compared to other types of data exfiltration. Additionally, due to the high adoption of the QUIC protocol in popular web services \cite{cloudflarehttp3}, a QUIC-based data exfiltration attack may achieve high throughput without raising suspicion.

\subsection{Additional Insights from Leading Firewall Vendors}

Firewall Vendor C mentioned that most requests regarding the QUIC protocol are coming from researchers, with only very little coming from industry. This indicates a divergence between academic interest in the protocol's potential and the industry's current level of adoption or need for it. Firewall Vendor D argues that the small performance gain through a reduced RTT does not justify the manageability challenges and potential security risks it entails. The vendors also correctly point out that tracking QUIC connections via a CID state table is not feasible, since the privacy-preserving mechanisms in QUIC may change CIDs at any time to prevent middleboxes from uniquely identifying a connection. A set of usable CIDs is negotiated as part of the encrypted QUIC handshake, and thus remains hidden from middleboxes.

\end{document}